\documentclass[twocolumn,prl, twocolumn,superscriptaddress, nofootinbib]{revtex4-1}

\usepackage[T1]{fontenc}
\usepackage[latin9]{inputenc}
\usepackage{color}
\usepackage{amsmath}
\usepackage{amssymb}
\usepackage{graphicx}
\usepackage[unicode=true,pdfusetitle,
 bookmarks=true,bookmarksnumbered=false,bookmarksopen=false,
 breaklinks=false,pdfborder={0 0 0},pdfborderstyle={},backref=false,colorlinks=true]
 {hyperref}
\hypersetup{
 citecolor=blue,urlcolor=blue}

\makeatletter

\newcommand{\lyxdot}{.}

\usepackage{bbm}
\usepackage[final]{changes}
\usepackage{ulem}

\makeatother

\begin{document}
\title{Controlling superconducting transistor by coherent light}
\author{Guo-Jian Qiao }
\affiliation{Beijing Computational Science Research Center, Beijing 100193, China}
\author{Zhi-Lei Zhang}
\affiliation{Graduate School of China Academy of Engineering Physics, Beijing 100193,
China}
\author{Sheng-Wen Li}
\email{lishengwen@bit.edu.cn}

\affiliation{Center for Quantum Technology Research, and Key Laboratory of Advanced
Optoelectronic Quantum Architecture and Measurements, School of Physics,
Beijing Institute of Technology, Beijing 100081, People\textquoteright s
Republic of China}
\author{C. P. Sun}
\email{suncp@gscaep.ac.cn}

\affiliation{Beijing Computational Science Research Center, Beijing 100193, China}
\affiliation{Graduate School of China Academy of Engineering Physics, Beijing 100193,
China}
\affiliation{School of Physics, Peking University, Beijing 100871, China}
\begin{abstract}
The Josephson junction is typically tuned by a magnetic field or electrostatic
gates to realize a superconducting transistor, which manipulates the
supercurrent in integrated superconducting circuits. However, this
tunable method does not achieve simultaneous control for the supercurrent
phase (phase difference between two superconductors) and magnitude.
Here, we propose a novel scheme for the light-controlled superconducting
transistor, which is composed of two superconductor leads linked by
a coherent light-driven quantum dot. We discover a Josephson-like
relation for supercurrent $I_{\mathrm{s}}=I_{c}(\Phi)\,\sin\Phi$,
where both supercurrent phase $\Phi$ and magnitude $I_{c}$ could
be entirely controlled by the phase, intensity, and detuning of the
driving light. Additionally, the supercurrent magnitude displays a
Fano profile with the increase of the driving light intensity, which
is clearly understood by comparing the level splitting of the quantum
dot under light driving and the superconducting gap. Moreover, when
two such superconducting transistors form a loop, they make up a light-controlled
superconducting quantum interference device (SQUID). Such a light-controlled
SQUID could demonstrate the Josephson diode effect, and the optimized
non-reciprocal efficiency achieves up to $54\%$, surpassing the maximum
record reported in recent literature. Thus, our feasible scheme delivers
a promising platform to perform diverse and flexible manipulations
in superconducting circuits.
\end{abstract}
\maketitle
\noindent\textbf{Introduction} - Semiconductor transistor lies in
the fundamental position in modern electronics. Analogous to it, superconducting
transistors, which can control superconducting current \citep{Ando2020_Observation_of_SDE,Daido_2022_Intrinsic_Superconducting_Diode,Baumgartner2022_Supercurrent_rectificatio,Bauriedl2022_SDE_magnetochiral_anisotropy,Wu2022_The_field_free_JD,Kouwenhoven_2006_Quantum_supercurrent_transistors,Kouwenhoven_2010_Hybrid_superconductor_QD},
are attracting more and more attentions, since they have lower power
consumption, and may provide potential applications for quantum information
processing \citep{Clarke2008_Superconducting_quantum_bits}. Recently,
superconducting transistors based on Josephson junction microstructures
\citep{Hu_2007_Proposed_Design_DJ,Nagaosa_2021_nonreciprocal_DE,Wu2022_The_field_free_JD}
and the superconducting quantum interferometer devices (SQUID) \citep{Szombati2016_phi_junction,Mikhailov_2022_SQUID_as_JD,Souto_2022_JDE_in_Supercurrent_Interferometers}
already have implemented to rectify (significant non-reciprocal supercurrent),
switch \citep{Yong_2005_Tunable_Supercurren,Kouwenhoven_2006_Quantum_supercurrent_transistors,Winkelmann2009_C_transistor,Katsaros_2010_Hybrid_S_semiconductor_devices,Silaev_2021_S_F_S_JJ}
and reverse \citep{baselmansReversingDirectionSupercurrent1999,Guinea_2001_J_coupling_through_QD,Marcus_2020_QD_Parity_JJ}
the supercurrent \citep{Kouwenhoven_2006_Supercurrent_reversal,Kouwenhoven_2010_Hybrid_superconductor_QD,Paolucci_2019_Field_Effect_Controllable_JI}.
Usually, such devices need to be controlled by an external magnetic
field or/and electrostatic voltage, to our knowledge, which are the
only two kinds of available control methods thus far \citep{Kouwenhoven_2006_Quantum_supercurrent_transistors,Baumgartner2022_Supercurrent_rectificatio,Fu_2022_Universal_SDE,Jiang_2022_JD_General_Theory}.
However, a concise scheme of the superconducting transistor based
on the Josephson junction by the above two tunable methods to simultaneously
control the supercurrent switch, reverse, phase (the phase difference
between two superconductors), and magnitude (critical current), has
yet to be proposed.

\begin{figure}
\includegraphics[width=1\columnwidth]{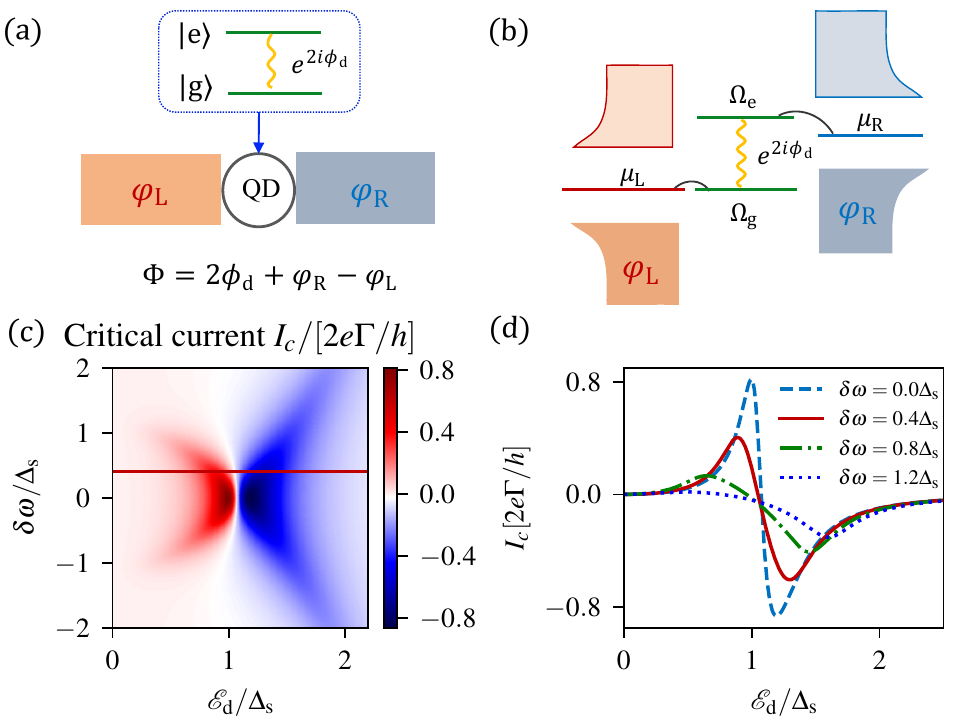}

\caption{(a) Two superconductors are linked by a light-driven quantum dot with
two levels, whose levels are illustrated in (b). (c) The change of
critical current $I_{c}$ with the the driving strength $\mathcal{E}_{\mathrm{d}}$
and the light detuning $\delta\omega$. (d) The critical current $I_{c}$
as the driving strength $\mathcal{E}_{\mathrm{d}}$ increase shows
the Fano profile with the fixed light detuning $\mathrm{\delta\omega=0,0.4,0.8,1.2\Delta_{\mathrm{s}}}$.
Hereafter, the same coupling strength always are set as $\Gamma_{\mathrm{L}}=\Gamma_{\mathrm{R}}=\Gamma=0.1\Delta_{\mathrm{s}}$.}

\label{Fig:phase-dependont-on-light}
\end{figure}

In this letter, we propose a novel scheme for the superconducting
transistor which is coherently controlled by light. The system is
composed of two superconductor leads linked by a light-driven quantum
dot (QD) with two levels {[}Fig. \ref{Fig:phase-dependont-on-light}
(a){]}, which is similar to a single-photon transistor based on superconducting
transmission line resonators and qubits \citep{Zhou_Controllable_Scattering_2008}.
When there is no driving light, the supercurrent is switched off,
while it is switched on under proper driving light intensity. More
importantly, it turns out the supercurrent passing through the transistor
depends not only on the driving light intensity, but also on the phase
$\phi_{\mathrm{d}}$ of the coherent light, as well as the superconductor
phase difference $\varDelta\varphi_{\mathrm{s}}$ between the two
leads, which gives $I_{\mathrm{s}}=I_{c}(\Phi)\,\sin\Phi$, with $\Phi:=\varDelta\varphi_{\mathrm{s}}+2\phi_{\mathrm{d}}$.
In this sense, we refer to it as the light-controlled dc Josephson
effect.

Moreover, it is intriguing that the critical current $I_{c}$ achieve
complete reversal by adjusting the detuning and driving intensity
of light, which implements a light-controlled $\pi$-junction in Fig.
\ref{Fig:phase-dependont-on-light} (c) \citep{baselmansReversingDirectionSupercurrent1999,Guinea_2001_J_coupling_through_QD,Tsung_han_2002_Microwave-induced,Kouwenhoven_2006_Supercurrent_reversal,Marcus_2020_QD_Parity_JJ,Marcus_2021_Zeeman_driven_parity_transitions,Christian2022_atomic_scaleSQI,Bargerbos_2022_Singlet_Doublet_Transitions_Quantum_Dot_Josephson_Junction}.
Remarkably, the critical current $I_{c}$ here exhibits a Fano dependence
on the driving light intensity with the fixed light detuning {[}see
Fig. \ref{Fig:phase-dependont-on-light} (d){]}. Namely, with the
increase of the driving light intensity, the critical current $I_{c}$
first increases in the positive direction, but then decreases down
to zero and increases in the negative direction, finally $I_{c}$
drops back and approaches zero. From the resonance comparison between
the QD levels subjected to driving light and the superconductor band
spectrum, we obtain a clear picture to understand this result. 

Further, an interference loop is formed by such two superconducting
transistors, and the light fields applied on the two QDs generate
a phase difference depending on their spatial distance. As a result,
the supercurrents passing through such an interference loop forward
and backward turn out to be asymmetric with each other, and a light-controlled
Josephson diode is achieved. It turns out the optimal nonreciprocity
of such a Josephson diode reaches $54\%$ by properly adjusting the
light intensity, which exceeds the best record reported up to now
($\sim40\%$) \citep{Fu_2022_Universal_SDE,Souto_2022_JDE_in_Supercurrent_Interferometers}. 

\vspace{0.6em}\noindent\textbf{Superconducting transistor setup}
- The setup of the superconducting transistor is demonstrated in Fig.\,\ref{Fig:phase-dependont-on-light}
(a), which is composed of a two-level QD in contact with two \textsl{s}-wave
superconductor leads. Denoting $\Delta_{\mathrm{n}}\equiv|\Delta_{\mathrm{n}}|e^{i\varphi_{\mathrm{n}}}$
and $\mu_{\mathrm{n}}$ as the pairing strength and chemical potential
of the two leads $\mathrm{n}=\mathrm{L},\mathrm{R}$, the Hamiltonian
of lead-$\mathrm{n}$ is described by \citep{Levy_1996_Hamiltonian_approach_to_transport,Lin_2000_Electron_transport_through_mesoscopic_hybrid}
\begin{equation}
H_{\mathrm{n}}=\sum_{\mathbf{k\sigma}}(\epsilon_{\mathrm{n},\mathbf{k}}+\mu_{\mathrm{n}})c_{\mathrm{n},\mathbf{k}\sigma}^{\dagger}c_{\mathrm{n},\mathbf{k}\sigma}+[e^{2i\mu_{\mathrm{n}}t}\Delta_{\mathrm{n}}^{*}c_{\mathrm{n},-\mathbf{k}\downarrow}c_{\mathrm{n},\mathbf{k}\uparrow}+\mathrm{H.c.}]\label{eq:Bcs-H}
\end{equation}
 with the index for spin $\sigma=\uparrow,\,\downarrow$. Hereafter,
we consider the two leads have $|\Delta_{\mathrm{R}}|=|\Delta_{\mathrm{L}}|:=\Delta_{\mathrm{s}}$,
but different phases $\varphi_{\mathrm{R,L}}$ and chemical potentials
$\mu_{\mathrm{R,L}}$.

The two leads are linked by a QD with two fermionic levels \citep{Recher_2010_Light_Emitting_Diode,Katsaros_2010_Hybrid_S_semiconductor_devices},
which is described by $H_{0}=\sum_{\sigma}\,\Omega_{\mathrm{e}}\,d_{\mathrm{e}\sigma}^{\dagger}d_{\mathrm{e}\sigma}+\Omega_{\mathrm{g}}\,d_{\mathrm{g}\sigma}^{\dagger}d_{\mathrm{g}\sigma}$
($\Omega_{\mathrm{e}}\ge\Omega_{\mathrm{g}}$). And, the electrons
tunnel between the QD and the two leads via the interaction 
\begin{equation}
H_{t}=-\sum_{\mathrm{i}=\mathrm{g,e}}\sum_{\mathrm{n},\mathbf{k},\sigma}\big(\mathrm{t}_{\mathrm{n,i},\mathbf{k}}d_{\mathrm{i}\sigma}^{\dagger}c_{\mathrm{n},\mathbf{k}\sigma}+\mathrm{t}_{\mathrm{n,i},\mathbf{k}}^{*}c_{\mathrm{n},\mathbf{k}\sigma}^{\dagger}d_{\mathrm{i}\sigma}\big).\label{eq:tunneling-L-Q-R}
\end{equation}
In the interaction picture, the tunneling terms between the upper
(lower) level and lead-L(R) would contain a fast oscillating factor,
and they could be omitted by the rotating wave approximation {[}See
Appendix \ref{sec:DcJE-Hamiltonian}{]}. Thus, equivalently the upper
(lower) level is only contacted with the right (left) lead {[}see
Fig. \ref{Fig:phase-dependont-on-light} (b){]}. 

Because of the separation of the QD, no supercurrent could flow between
the two superconducting leads. However, a tunneling bridge could be
built up when a monochromatic coherent light (with frequency $\omega_{\mathrm{d}}$)
is injected on the QD, which drives the electron up and down between
the two QD levels: $V_{\mathrm{d}}=\sum_{\sigma}\,\mathcal{E}_{\mathrm{d}}\,d_{\mathrm{e}\sigma}^{\dagger}d_{\mathrm{g}\sigma}e^{-i\omega_{\mathrm{d}}t-i\phi_{\mathrm{d}}}+\mathrm{H.c.}$
Here, $\phi_{\mathrm{d}}$ comes from the phase of the coherent light,
$\mathcal{E}_{\mathrm{d}}$ is taken as a positive and real driving
strength, and $|\mathcal{E}_{\mathrm{d}}|^{2}$ is proportional to
the light intensity. 

\vspace{0.6em}\noindent\textbf{Light-controlled dc Josephson effect}
- In the rotating frame applied by the unitary transformation $\mathbf{U}(t)=\exp[i(N_{\mathrm{L}}+\sum_{\sigma}d_{\mathrm{g}\sigma}^{\dagger}d_{\mathrm{g}\sigma})\mu_{\mathrm{L}}t+i(N_{\mathrm{R}}+\sum_{\sigma}d_{\mathrm{e}\sigma}^{\dagger}d_{\mathrm{e}\sigma})(\mu_{\mathrm{R}}t+\phi_{\mathrm{d}})]$
with $N_{\mathrm{n}}=\sum_{\mathbf{k\sigma}}c_{\mathrm{n},\mathbf{k}\sigma}^{\dagger}c_{\mathrm{n},\mathbf{k}\sigma}$,
the Hamiltonian of the QD under coherent driving becomes 
\begin{equation}
\bar{H}_{\text{\textsc{qd}}}=\sum_{\mathrm{i,}\sigma}\bar{\Omega}_{\mathrm{i}}d_{\mathrm{i}\sigma}^{\dagger}d_{\mathrm{i}\sigma}-(\mathcal{E}_{\mathrm{d}}d_{\mathrm{e}\sigma}^{\dagger}d_{\mathrm{g}\sigma}e^{i(\mu_{\mathrm{R}}-\mu_{\mathrm{L}}-\omega_{\mathrm{d}})t}+\mathrm{H.c.})\label{eq:Light-driven-quantum-dot-H}
\end{equation}
with the reduced upper (lower) level $\bar{\Omega}_{\mathrm{e}(\mathrm{g})}:=\Omega_{\mathrm{e}(\mathrm{g})}-\mu_{\mathrm{R}(\mathrm{L})}$.

\begin{figure}
\includegraphics[width=1\columnwidth]{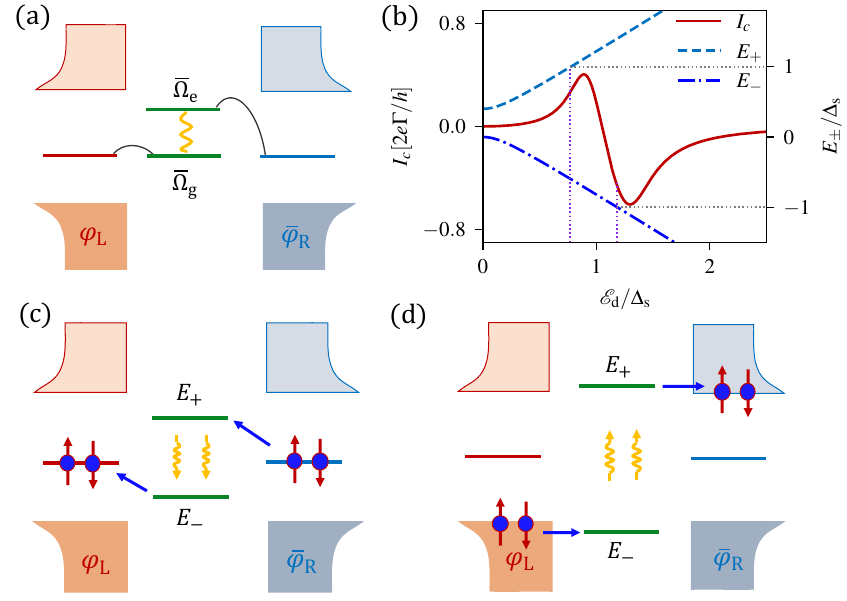}

\caption{(a) The level distribution of the superconducting transistor setup
in the rotating frame. (b) The split of energy levels $E_{+}$ (dashed
line) and $E_{-}$ (dashdotted line) of QD with the driving strength
$\mathcal{E}_{\mathrm{d}}$ increase. The intersections of the purple
and gray dotted lines corresponds to $E_{\pm}=\Delta_{\mathrm{s}}$.
The Fano profile of critical current is because the flow direction
of Cooper pairs reverses the back {[}from lead-R to lead-L in (c){]}
and forth {[}from lead-L to lead-R in (d){]} between two superconducting
leads with increasing driving strength.}

\label{Fig:Fano_shape}
\end{figure}

Meanwhile, in this rotating frame, the Hamiltonian (\ref{eq:Bcs-H})
for the two leads become time-independent, meaning that the effective
chemical potentials of two superconducting leads are regarded as $\bar{\mu}_{\mathrm{L}}=\bar{\mu}_{\mathrm{R}}=0$
, and the phases of the two superconducting leads are also corrected
as $\bar{\varphi}_{\mathrm{R}}=\varphi_{\mathrm{R}}+2\phi_{\mathrm{d}}$
and $\varphi_{\mathrm{L}}$ {[}see Fig.\,\ref{Fig:Fano_shape}(a){]}.
Further, if we focus on the situation that $\omega_{\mathrm{d}}=\mu_{\mathrm{R}}-\mu_{\mathrm{L}}$,
namely, the driving frequency is equal to the voltage difference,
and thus the Hamiltonian (\ref{eq:Light-driven-quantum-dot-H}) also
becomes time-independent.

Based on this time-independent model in the rotating frame, by calculating
the evolution of the Green functions of this system, the supercurrent
through the superconducting transistor induced by the driving light
is exactly obtained as
\begin{equation}
I_{\mathrm{s}}=I_{c}(\Phi)\sin(\Phi),\label{eq:Dc-current}
\end{equation}
where the verbose critical current $I_{c}$ is explicitly given in
Appendix \ref{sec:DcJE-Hamiltonian}. Especially, here $\Phi$ is
a phase difference defined by 
\begin{equation}
\Phi:=\bar{\varphi}_{\mathrm{R}}-\varphi_{\mathrm{L}}=\varphi_{\mathrm{R}}-\varphi_{\mathrm{L}}+2\phi_{\mathrm{d}}.\label{eq:total-phase}
\end{equation}
It is worth noting that the above supercurrent follows the Josephson-like
relation, but here $\Phi$ incorporates not only the phase difference
from the two superconducting leads $\varDelta\varphi_{\mathrm{s}}\equiv\varphi_{\mathrm{R}}-\varphi_{\mathrm{L}}$,
but also an additional correction from the light phase $\phi_{\mathrm{d}}$
which was previously attainable only by a magnetic field passing through
the junction \citep{tinkham2004introduction}. Hereafter, we set $\mu_{\mathrm{R}}\ge\mu_{\mathrm{L}}=\Omega_{\mathrm{g}}\equiv0$
as the energy reference and define the detuning between the driving
frequency $\omega_{\mathrm{d}}$ and the energy gap of the two QD
levels (light detuning) as $\delta\omega=(\Omega_{\mathrm{e}}-\Omega_{\mathrm{g}})-\omega_{\mathrm{d}}$. 

In the large detuning situation ($\delta\omega\gg\mathcal{E}_{\mathrm{d}}$),
the above critical current $I_{c}(\Phi)\simeq I_{c}$ becomes a constant
independent of the phase difference $\Phi$, then the above supercurrent
(\ref{eq:Dc-current}) is simplified as $I_{\mathrm{s}}=I_{c}\sin\Phi$
{[}see Appendix \ref{sec:DcJE-Hamiltonian}{]}. This result well returns
the same form as the dc Josephson effect \citep{josephson1962possible},
except here $\Phi$ is corrected by the light phase $\phi_{\mathrm{d}}$. 

On the other hand, it is evident that the critical current $I_{c}(\Phi)$
is also highly dependent on the driving strength $\mathcal{E}_{\mathrm{d}}$,
or the driving light intensity. When there is no driving light ($\mathcal{E}_{\mathrm{d}}=0$),
no supercurrent flows across the two leads $I_{\mathrm{s}}=0$, which
is consistent with our previous discussion. As displayed in Fig. \ref{Fig:Fano_shape}
(b), the critical current $I_{c}$ depending on the driving strength
$\mathcal{E}_{\mathrm{d}}$ as well as the detuning $\delta\omega$
(here the total phase difference is set as $\Phi=\pi/2$) appears
as a butterfly-like shape. Notably, when the detuning is fixed as
$\delta\omega=0.4\Delta_{\mathrm{s}}$ {[}solid red line of Fig. \ref{Fig:phase-dependont-on-light}
(c){]}, the critical current $I_{c}$ exhibits a Fano profile depending
on the driving strength $\mathcal{E}_{\mathrm{d}}$ {[}Fig. \ref{Fig:phase-dependont-on-light}
(d){]}, which could be fully reversed in both positive and negative
directions. In this case, this superconducting transistor achieves
a  $\pi$-junction controlled by the light intensity \citep{baselmansReversingDirectionSupercurrent1999,Guinea_2001_J_coupling_through_QD,Tsung_han_2002_Microwave-induced,Kouwenhoven_2006_Supercurrent_reversal,Marcus_2020_QD_Parity_JJ,Marcus_2021_Zeeman_driven_parity_transitions,Christian2022_atomic_scaleSQI,Bargerbos_2022_Singlet_Doublet_Transitions_Quantum_Dot_Josephson_Junction}. 

\vspace{0.6em}\noindent\textbf{Fano dependence of the critical current}
- Unexpectedly, as the driving light increases, the critical current
does not increase, but exhibits a Fano profile as depicted in Fig.
\ref{Fig:Fano_shape} (b), which has been observed in controllable
single-photon transistors as well \citep{Zhou_Controllable_Scattering_2008}.
And this Fano profile is elucidated by considering the weak coupling
limit $\Gamma\ll\Delta_{\mathrm{s}}$, which allows us to neglect
high order terms $\Gamma/\Delta_{\mathrm{s}}$. In this limit, the
critical current (\ref{eq:Dc-current}) is reduced to

\begin{equation}
I_{c}\simeq\int\mathrm{d}\nu\frac{e\Gamma\sqrt{\gamma_{\mathrm{s}}(\nu)}\Delta_{\mathrm{s}}^{2}\mathcal{E}_{\mathrm{d}}^{2}}{h\left[2\nu^{2}-\Omega^{2}\right]\nu^{2}}\frac{\Gamma^{2}}{\Gamma^{2}+\frac{\gamma_{\mathrm{s}}^{2}(\nu)(\nu^{2}-E_{+}^{2})^{2}(\nu^{2}-E_{-}^{2})^{2}}{\nu^{4}\left((\nu^{2}-E_{+}^{2})+(\nu^{2}-E_{-}^{2})\right)^{2}}},\label{eq:simplied_DC_1}
\end{equation}
with $\gamma_{\mathrm{s}}(\nu)=\sqrt{\nu^{2}-\Delta_{\mathrm{s}}^{2}}$
and $\Omega^{2}=2(\delta\omega^{2}+2\mathcal{E}_{\mathrm{d}}^{2})$.
And, the QD levels are obtained by diagonalizing the Hamiltonian (\ref{eq:Light-driven-quantum-dot-H})
of light-driven QD as $E_{\pm}=(1/2)(\delta\omega\pm\sqrt{\delta\omega^{2}+4\mathcal{E}_{\mathrm{d}}^{2}})$.
By taking into account the main contribution of the integral (\ref{eq:simplied_DC_1})
near $\nu\simeq-\Delta_{\mathrm{s}}-\varepsilon$ ($\varepsilon$
is infinitesimal) due to $(\nu^{2}-\Delta^{2})^{-\frac{1}{2}}\rightarrow\infty$,
thus the critical current is approximated as 
\begin{equation}
I_{c}\simeq\frac{2e\Gamma}{h}\frac{\mathcal{E}_{\mathrm{d}}^{2}(2\varepsilon\Delta_{\mathrm{s}})^{-\frac{1}{2}}}{2\Delta_{\mathrm{s}}^{2}-\Omega^{2}}\frac{\Gamma^{2}}{\Gamma^{2}+\frac{2\varepsilon(\Delta_{\mathrm{s}}^{2}-E_{+}^{2})^{2}(\Delta_{\mathrm{s}}^{2}-E_{-}^{2})^{2}}{\Delta_{\mathrm{s}}^{3}\left((\Delta^{2}-E_{+}^{2})+(\Delta^{2}-E_{-}^{2})\right)^{2}}}.\label{eq:simplied_DC_2}
\end{equation}
And numerical computation reveals that the aforementioned approximation
is reasonable {[}see Appendix \ref{sec:DcJE-Hamiltonian}{]}. Notice
that, as the driving strength increase, the critical current $(\ref{eq:simplied_DC_2})$
appears with two resonance peaks corresponding to $E_{\pm}=\Delta_{\mathrm{s}}$,
the positive (negative) of which is determined by the sign of $2\Delta_{\mathrm{s}}^{2}-(\delta\omega^{2}+2\mathcal{E}_{\mathrm{d}}^{2})$.

Specifically, when the driving light intensity is weak, the two QD
levels are positioned beneath the superconducting gap {[}Fig. \ref{Fig:Fano_shape}
(c){]}, effectively creating Andreev bound states \citep{zagoskin1998quantum}.
These two states serves as a pathway for supercurrent transport, whereby
the Cooper pair flows from the lead-R into the upper QD level through
the Andreev reflection (AR), then transitions to the lower QD level
via light emission, and finally flows into the lead-L by AR again
{[}Fig. \ref{Fig:Fano_shape} (c){]}. Meanwhile, the critical current
is positive due to $2\Delta_{\mathrm{s}}^{2}-(\delta\omega^{2}+2\mathcal{E}_{\mathrm{d}}^{2})>0$. 

As the intensity of the driving light further increases, one of the
two QD levels will be adjacent to the superconducting continuum spectrum,
resulting in the emergence of a resonant peak of positive critical
current. However, when one of two QD levels surpasses the superconducting
gap, the positive critical current gradually decreases zero at $\mathcal{E}_{\mathrm{d}}^{2}=\Delta_{\mathrm{s}}^{2}-\delta\omega^{2}/2$
and even reverse as AR is further suppressed. Once both QD levels
surpass the superconducting gap {[}Fig. \ref{Fig:Fano_shape} (d){]},
the supercurrent increases negatively, and another resonance peak
of negative critical current appears. At this point, Cooper pair flows
from lead-L to the lower QD level, then driven by light to the upper
QD level and finally transports into the continuous band of lead-R
{[}Fig. \ref{Fig:Fano_shape} (d){]}. 

The above physical picture have also been justified. As shown in Fig.
\ref{Fig:Fano_shape} (b), the driving strength $\mathcal{E}_{\mathrm{d}}^{\pm}$
corresponding to the intersection point of $E_{\pm}(\mathcal{E}_{\mathrm{d}})=\pm\Delta_{\mathrm{s}}$
{[}two dotted purple lines in Fig. \ref{Fig:Fano_shape} (b){]} and
the position $\mathcal{E}_{\mathrm{peak}}^{\pm}$ of two resonant
peaks of the critical current $I_{c}(\mathcal{E}_{\mathrm{peak}})$
are approximately equal within the range of $\Gamma$ error, i.e.,
$|\mathcal{E}_{\mathrm{d}}^{\pm}-\mathcal{E}_{\mathrm{peak}}^{\pm}|\leq\Gamma$.
And the deviation of the approximate theory analysis using Eq. (\ref{eq:simplied_DC_2})
and the accurate numerical computation is ascribed to the application
of the weak coupling condition $\Gamma\ll\Delta_{\mathrm{s}}$ and
exclusive consideration of the primary contribution of the integral
near $\nu\sim-\Delta_{\mathrm{s}}$.

\begin{figure}
\includegraphics[width=1\columnwidth]{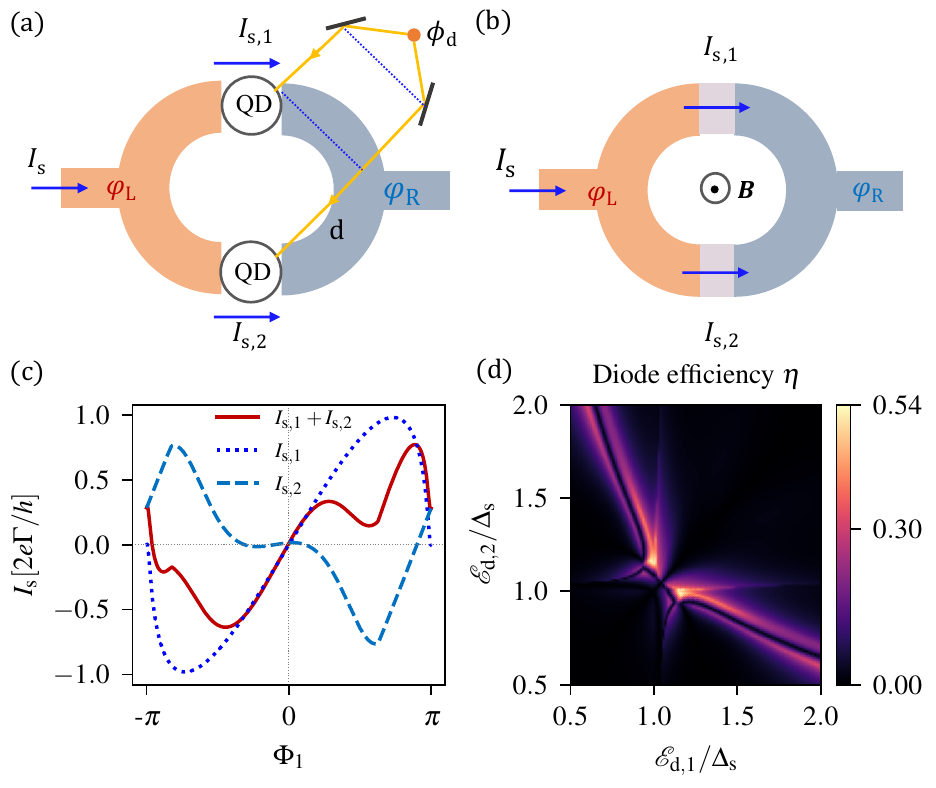}

\caption{The supercurrent through superconducting quantum interference device
controlled by the coherent light (a) and the magnetic field (b). (c)
The total phase-asymmetric current (red solid line) is sum of the
phase-symmetric current (blue dotted line) through QD-$1$ and the
phase-symmetric current (light blue dashed line) through QD-$2$ with
the driving strength $\mathcal{E}_{\mathrm{d},1}=1.0\Delta_{\mathrm{s}}$
and $\mathcal{E}_{\mathrm{d},2}=1.1\Delta_{\mathrm{s}}$. (d) Diode
efficiency $\eta$ versus driving strength $\mathcal{E}_{\mathrm{d},1}$
and $\mathcal{E}_{\mathrm{d},2}$ for QD-$1$ and QD-$2$. The optical
phase difference and light detuning have set: $\varDelta\phi_{\mathrm{d}}=\pi/10$
and $\delta\omega=0$.}

\label{Fig:SQUID}
\end{figure}

\vspace{0.6em}\noindent\textbf{Light-controlled SQUID} - Furthermore,
if two such superconducting transistors are placed nearby and forms
a loop, they make up a SQUID which is also coherently controlled by
light \citep{Recher_2010_Light_Emitting_Diode,Yuli_2014_Light_superconducting_interference_devices,Bouscher_2017_Sem_Sc_opto_devices}
{[}Fig. \ref{Fig:SQUID} (a){]}. The current through the SQUID sums
up the contributions through each QD, $I_{\mathrm{s}}=I_{\mathrm{s},1}+I_{\mathrm{s},2}$,
where the current $I_{\mathrm{s},i}$ through QD-$i$ is given by
the modified Josephson relation (\ref{eq:Dc-current}). Under a proper
driving light intensity ($\mathcal{E}_{\mathrm{d}}\ll\delta\omega$),
the current through the SQUID gives {[}see Appendix \ref{sec:current-SQUID}{]}
\begin{equation}
I_{\mathrm{s}}\simeq I_{c,1}\sin(\varDelta\varphi_{\mathrm{s}}+2\phi_{\mathrm{d},1})+I_{c,2}\sin(\varDelta\varphi_{\mathrm{s}}+2\phi_{\mathrm{d},2}),\label{eq:Op-SQUID}
\end{equation}
with the phase difference between two superconductors $\varDelta\varphi_{\mathrm{s}}\equiv\varphi_{\mathrm{R}}-\varphi_{\mathrm{L}}$
and the phase $\phi_{\mathrm{d},i}$ of driving light applied to QD-$i$.
Here we consider the light fields applied on the two QDs originate
from the same point source, thus they have a deterministic phase difference
which is determined by their optical paths $\varDelta\phi_{\mathrm{d}}\equiv\phi_{\mathrm{d},2}-\phi_{\mathrm{d},1}$
{[}see Fig. \ref{Fig:SQUID}(a){]}.

If the light intensities applied on the two QDs are equal, they give
equal critical currents $I_{c,1}=I_{c,2}\equiv I_{c}$, and then the
supercurrent through the SQUID becomes $I_{\mathrm{s}}=2I_{c}\sin(\varDelta\varphi_{\mathrm{s}}+\phi_{\mathrm{d},1}+\phi_{\mathrm{d},2})\cos\varDelta\phi_{\mathrm{d}}$.
This is similar to the SQUID controlled by the magnetic flux {[}Fig.
\ref{Fig:SQUID} (b){]} \citep{tinkham2004introduction}. The sinusoidal
oscillation of the supercurrent adjusted by changing the optical path
difference provides a method to set a comparison reference between
the superconducting phase and the light phase.

More generally, if the light intensities applied on the two QDs are
not equal to each other and generate different critical currents $I_{c,1}(\mathcal{E}_{\mathrm{d},1},\phi_{\mathrm{d},1})\neq I_{c,2}(\mathcal{E}_{\mathrm{d},2},\phi_{\mathrm{d},2})$,
it turns out such a SQUID gives an asymmetric current-phase relation
\citep{Golubov_2004_current_phase_relation}, which can be utilized
to realize a Josephson diode \textbf{\citep{Souto_2022_JDE_in_Supercurrent_Interferometers}}.
From the Josephson relation (\ref{eq:Dc-current}), it can be verified
that the current through QD-$1$ $I_{\mathrm{s},1}=I_{c,1}(\Phi_{1})\sin\Phi_{1}$
is an odd function of $\Phi_{1}=\varDelta\varphi_{\mathrm{s}}+2\phi_{\mathrm{d},1}$,
since $I_{c,1}(\Phi_{1})=I_{c,1}(-\Phi_{1})$ is even; in contrast,
because of the optical phase difference $\varDelta\phi_{\mathrm{d}}$,
the current through QD-2 $I_{\mathrm{s},2}(\Phi_{2}\equiv\Phi_{1}+2\varDelta\phi_{\mathrm{d}})$
is asymmetric of $\Phi_{1}$ {[}see Fig.\,\ref{Fig:SQUID}(c){]}. 

As a result, summing up $I_{\mathrm{s},1}$ and $I_{\mathrm{s},2}$,
the maximum currents through the SQUID in the two directions are different,
$|I_{\mathrm{s},max}^{+}|\neq|I_{\mathrm{s},max}^{-}|$ {[}Fig.\,\ref{Fig:SQUID}
(c){]}, which makes the SQUID a diode. To quantify the non-reciprocity
of the diode, we adopt the following diode efficiency \citep{Fu_2022_Universal_SDE,Souto_2022_JDE_in_Supercurrent_Interferometers}
\begin{equation}
\eta=\frac{\left||I_{\mathrm{s},max}^{+}|-|I_{\mathrm{s},max}^{-}|\right|}{|I_{\mathrm{s},max}^{+}|+|I_{\mathrm{s},max}^{-}|}.
\end{equation}
We find that, when the optical path difference is set as $\varDelta\phi_{\mathrm{d}}\simeq0.1\pi$,
and the driving strengths applied on the two QDs are set as $\mathcal{E}_{\mathrm{d},1}\simeq0.99\Delta_{\mathrm{s}}$,
$\mathcal{E}_{\mathrm{d},2}\simeq1.16\Delta_{\mathrm{s}}$, the above
diode efficiency achieves its maximum $\eta\simeq53.9\%$ {[}see Fig.\,\ref{Fig:SQUID}
(d){]}. This is greater than the best record reported in literature
($\sim40\%$), which was controlled by magnetic field \citep{Fu_2022_Universal_SDE,Souto_2022_JDE_in_Supercurrent_Interferometers}. 

\vspace{0.6em}\noindent\textbf{Conclusion} - We propose a novel scheme
for a coherently controlled superconducting transistor that operates
under a properly driven light. The supercurrent through this transistor
adheres to a Josephson-like relation $I_{\mathrm{s}}=I_{c}(\Phi)\,\sin\Phi$,
but it is worth noting that the phase $\Phi=\Delta\varphi_{\mathrm{s}}+2\phi_{\mathrm{d}}$
here incorporates not only the phase difference between the two superconducting
leads $\Delta\varphi_{\mathrm{s}}$, but also the phase of the driving
light $\phi_{\mathrm{d}}$. The critical current $I_{c}$ in our Josephson
relation hinges on the phase, detuning and intensity of driving light,
and it shows a Fano profile with the increase of the driving light
intensity, which is elucidated by comparing the QD levels under light
driving and the superconducting gap. 

Furthermore, a light-controlled SQUID could be implemented by using
two such superconducting transistors. It turns out such a light-controlled
SQUID could realize the Josephson diode effect, and the optimized
non-reciprocal efficiency achieves up to $\sim54\%$, which is superior
to the maximum record reported in the recent literature. In this sense,
our new scheme provide a promising platform to achieve flexible and
diverse manipulations of supercurrent in superconducting circuits.

\vspace{0.6em}

This study is supported by NSF of China (Grant No.\,12088101 and
No.\,11905007), and NSAF (Grants No.\,U1930403 and No.\,U1930402).

\appendix
\begin{widetext}

\section{The generalied Landauer formula\label{sec:Landauer-formula}}

Usually, the current measurement of the system is achieved by connecting
with the local electron leads. Here, we consider the Hamiltonian of
the measured system as the quadratic fermions
\begin{equation}
H_{s}=\frac{1}{2}\mathbf{d}^{\dagger}\cdot\mathbf{H}\cdot\mathbf{d},\label{eq:measured-system}
\end{equation}
where the vector operation is $\mathbf{d}=[\mathsf{d}_{\mathrm{1}},\mathsf{d}_{\mathrm{2}},\dots,\mathsf{d}_{\mathrm{N}}]^{T}$
with the site-$\mathrm{n}$ electron operator $\mathsf{d}_{\mathrm{n}}=[d_{\mathrm{n}\uparrow},d_{\mathrm{n}\downarrow},d_{\mathrm{n}\uparrow}^{\dagger},d_{\mathrm{n}\downarrow}^{\dagger}]$
in Nambu notation. The site-$\mathrm{n}$ of the system is connected
with the electron reservoir $\alpha$, which is described by
\begin{equation}
H_{\mathrm{n}}^{\alpha}=\frac{1}{2}\sum_{\mathbf{k}}\mathbf{c}_{\mathrm{n},\mathbf{k}}^{\dagger}\cdot\mathsf{\mathbf{H}}_{\mathrm{n},\mathbf{k}}^{\alpha}\cdot\mathbf{c}_{\mathrm{n},\mathbf{k}}\label{eq:electron-lead}
\end{equation}
with the vector operation of the electron lead $\mathbf{c}_{\mathrm{n},\mathbf{k}}=(c_{\mathrm{n},\mathbf{k}\uparrow},c_{\mathrm{n},-\mathbf{k}\downarrow},c_{\mathrm{n},\mathbf{k}\uparrow}^{\dagger},c_{\mathrm{n},-\mathbf{k}\downarrow}^{\dagger})^{T}$.
The electron reservoir can be an \textsl{s}-wave superconductor, regarded
as the measured lead or providing the superconducting proximity effect,
and the Hamiltonian matrix for $\alpha=s$ is
\begin{equation}
\mathbf{H}_{\mathbf{\mathrm{n},\mathbf{k}}}^{\text{s}}:=\begin{bmatrix}\epsilon_{\mathrm{n},\mathbf{k}} &  &  & \Delta_{\mathrm{n}}\\
 & \epsilon_{\mathrm{n},\mathbf{k}} & -\Delta_{\mathrm{n}}\\
 & -\Delta_{\mathrm{n}}^{*} & -\epsilon_{\mathrm{n},\mathbf{k}}\\
\Delta_{\mathrm{n}}^{*} &  &  & -\epsilon_{\mathrm{n},\mathbf{k}}
\end{bmatrix}.\label{eq:lead-Matrix}
\end{equation}
When the pairing strength is zero, i.e. $\Delta_{\mathrm{n}}=0$,
the Hamiltonian matrix of metal leads $\alpha=e$ is obtained as $\mathbf{H}_{\mathrm{n},\mathbf{k}}^{\text{e}}=\text{diag}\{\epsilon_{\mathrm{n},\mathbf{k}},\epsilon_{\mathrm{n},\mathbf{k}},-\epsilon_{\mathrm{n},\mathbf{k}},-\epsilon_{\mathrm{n},\mathbf{k}}\}$.
The single electron tunneling interaction between site-$\mathrm{n}$
of the system and the lead is written as 
\begin{equation}
H_{t}=-\sum_{\mathbf{k}\sigma}(\mathrm{t}_{\mathrm{n},\mathbf{k}}\,d_{\mathrm{n},\sigma}^{\dagger}c_{\mathrm{n},\mathbf{k}\sigma}+\mathrm{H.c.}),\label{eq:tunneling}
\end{equation}

The current operation for the time $t$ is defined by the change of
total electron number $N_{\mathrm{n}}=\sum_{\mathbf{k}\sigma}c_{\mathrm{n},\mathbf{k}\sigma}^{\dagger}c_{\mathrm{n},\mathbf{k}\sigma}$
in the connected reservoir $\mathrm{n}$ as 
\begin{equation}
I_{\mathrm{n}}(t)=-2\frac{e}{\hbar}\sum_{\mathbf{k}}\Im[\langle\mathbf{d}^{\dagger}(t)\cdot\mathbf{P}^{+}\cdot\mathbf{T}_{\mathrm{n,}\mathbf{k}}\cdot\mathbf{c}_{\mathrm{n},\mathbf{k}}(t)\rangle]\label{eq:current-t}
\end{equation}
Here, the projection operator is $\mathbf{P}^{+}=\text{diag}\{\mathsf{P}^{+},\mathsf{P}^{+},\ldots,\mathsf{P}^{+}\}$
with $4\times4$ diagonal matrix $\mathsf{P}^{+}=\text{diag}\{1,1,0,0\}$,
and $\mathbf{T}_{\mathrm{n,}\mathbf{k}}=\big[0,\ldots,\mathsf{T}_{\mathrm{n},\mathbf{k}},\dots,\mathbf{0}\big]^{T}$
are the the $4N\times4$ tunneling matrices with $4\times4$ blocks
$\mathsf{T}_{\mathrm{n},\mathbf{k}}:=\mathrm{diag}\big\{\mathrm{t}_{\mathrm{n},\mathbf{k}},\mathrm{t}_{\mathrm{n},\mathbf{k}},-\mathrm{t}_{\mathrm{n},\mathbf{-k}}^{*}-\mathrm{t}_{\mathrm{n},\mathbf{-k}}^{*}\big\}$,
which indicates the coupling between the system and the reservoir
$\mathrm{n}$. The current (\ref{eq:current-t}) is obtained by the
time evolution of the Green's function $\partial_{t}\boldsymbol{G}_{ij}(t,t^{\prime})=i\langle\partial_{t}[\mathbf{d}^{\dagger}(t)]_{i}[\mathbf{c}_{\mathrm{n},\mathbf{k}}(t^{\prime})]_{j}\rangle$
with $t=t^{\prime}$, which is equivalent to calculating the time
evolution of the vector operator $\mathbf{d}(t)$ and $\mathbf{c}_{\mathbf{\mathrm{n,}\mathbf{k}}}(t)$
\citep{Meir_1992_Landauer_formula,Flensberg_2010_Tunneling_of_MBS}.
Then, it follows from the Heisenberg equation that the operator's
time evolution is obtained as 
\begin{equation}
\begin{aligned}\partial_{t}\mathbf{d}(t) & =\delta(t)\mathbf{d}(0)-i\mathbf{H}\cdot\mathbf{d}(t)+i\sum_{\mathrm{m}}\sum_{\mathbf{k}}\mathbf{T}_{\mathrm{m,}\mathbf{k}}\cdot\mathbf{c}_{\mathrm{m,}\mathbf{k}}(t),\\
\partial_{t}\mathbf{c}_{\mathbf{\mathrm{n,}\mathbf{k}}}(t) & =\delta(t)\mathbf{c}_{\mathrm{n,}\mathbf{k}}(0)-i\mathbf{H}_{\mathrm{n,}\mathbf{k}}^{\alpha}\cdot\mathbf{c}_{\mathrm{n,}\mathbf{k}}(t)+i\mathbf{T}_{\mathrm{n,}\mathbf{k}}^{\dagger}\cdot\mathbf{d}(t).
\end{aligned}
\label{eq:motion-eq}
\end{equation}
Here, the summation $\mathrm{m}$ contains all site positions of the
system connected by electron lead. The above linear equations of the
vector operator $\mathbf{c}_{\mathbf{\mathrm{n,}\mathbf{k}}}(t)$
are solvable by the Fourier transform as follow
\begin{align}
\mathbf{c}_{\mathbf{\mathrm{n,}\mathbf{k}}}(\omega) & =\mathsf{G}_{\mathrm{n,}\mathbf{k}}^{\alpha}(\omega)\cdot[\mathbf{c}_{\mathbf{k}}(0)+i\mathbf{T}_{\mathrm{n,}\mathbf{k}}^{\dagger}\cdot\mathbf{d}(\omega)],\label{eq:lead-G}
\end{align}
where the Green function of the lead is $\mathsf{G}_{\mathrm{n,}\mathbf{k}}(\omega)=i[\omega^{+}-\mathbf{H}_{\mathrm{n,}\mathbf{k}}^{\alpha}]^{-1}$
with $\omega^{+}\equiv\omega+i\varepsilon$ ($\varepsilon$ is infinitesimal).
Similarly, the dynamic evolution of the system in $\omega$-space
is exactly solved as
\begin{equation}
\mathbf{d}(\omega)=\mathbf{G}(\omega)\cdot\big[\mathbf{d}(0)+i\boldsymbol{\xi}(\omega)\big],\label{eq:system-G}
\end{equation}
Here, the Green function of the system is defined as $\mathbf{G}(\omega)=i[\omega^{+}-\mathbf{H}+i\mathbf{D}(\omega)]^{-1}$.
The random force of all the connected electron lead in Fourier space
is 
\begin{equation}
\boldsymbol{\xi}(\omega)=\sum_{\mathrm{m}}\boldsymbol{\xi}_{\mathrm{m}}(\omega)=\sum_{\mathrm{m}}\sum_{\mathbf{k}}\mathbf{T}_{\mathrm{m},\mathbf{k}}\cdot\mathsf{G}_{\mathrm{m},\mathbf{k}}^{\alpha}(\omega)\cdot\mathbf{c}_{\mathrm{m},\mathbf{k}}(0),\label{eq:random-force}
\end{equation}
and the disappation kernal caused by the electron reservoir is
\begin{align}
\mathbf{D}(\omega) & =\sum_{\mathrm{m}}\mathbf{D}_{\mathrm{m}}(\omega)=\sum_{\mathbf{k}}\mathbf{T}_{\mathrm{m},\mathbf{k}}\cdot\mathsf{G}_{\mathrm{m},\mathbf{k}}^{\alpha}(\omega)\cdot\mathbf{T}_{\mathrm{m},\mathbf{k}}^{\dagger}\equiv\sum_{\mathrm{m}}\frac{1}{2}\boldsymbol{\Gamma}_{\mathrm{m}}(\omega)+i\mathbf{V}_{\mathrm{m}}(\omega)\label{eq:dissaption}
\end{align}

Under the local tunneling approximation of the electron exchange,
i.e. $\mathrm{t}_{\mathrm{n},\mathbf{k}}\mathrm{t}_{\mathrm{m},\mathbf{k}}^{*}\simeq\delta_{\mathrm{n,m}}|\mathrm{t}_{\mathrm{n},\mathbf{k}}|^{2}$,
the dissipation kernel of the site-$\mathrm{n}$ connected by the
lead is simplified as $\mathbf{D}_{\mathrm{n}}(\omega)=\text{diag}\{\mathbf{0},\ldots,\mathsf{D}_{\mathrm{n}}(\omega),\ldots,\mathbf{0}\}$
with $4\times4$ blocks $\mathsf{D}_{\mathrm{n}}(\omega)=\frac{1}{2}\mathsf{\Gamma}_{\mathrm{n}}(\omega)+i\mathsf{V}_{\mathrm{n}}(\omega)$,
where the real part $\mathsf{\Gamma}_{\mathrm{n}}(\omega):=\mathsf{\Gamma}_{\mathrm{n}}^{+}(\omega)+\mathsf{\Gamma}_{\mathrm{n}}^{-}(\omega)$
leads to dissipation and the imaginary part $\mathsf{V}_{\mathrm{n}}(\omega)$
provides an effective interaction for the site-$\mathrm{n}$ by the
superconducting proximity effect \citep{Qiao_2022}
\begin{equation}
\mathsf{\Gamma}_{\mathrm{n}}^{\pm}(\omega):=\pm\frac{\Theta(\pm\omega-|\Delta_{\mathrm{n}}|)\Gamma_{\mathrm{n}}}{\sqrt{\omega^{2}-|\Delta_{\mathrm{n}}|^{2}}}\Sigma_{\mathrm{n}}(\omega),\,\mathsf{V}_{\mathrm{n}}(\omega)=-\frac{1}{2}\frac{\Theta(|\Delta_{\mathrm{n}}|-|\omega|)\Gamma_{\mathrm{n}}}{\sqrt{|\Delta_{\mathrm{n}}|^{2}-\omega^{2}}}\Sigma_{\mathrm{n}}(\omega),\,\Sigma_{\mathrm{n}}(\omega)=\left(\begin{array}{cccc}
\omega & 0 & 0 & -\Delta_{\mathrm{n}}\\
0 & \omega & \Delta_{\mathrm{n}} & 0\\
0 & \Delta_{\mathrm{n}}^{*} & \omega & 0\\
-\Delta_{\mathrm{n}}^{*} & 0 & 0 & \omega
\end{array}\right).\label{eq:disspation}
\end{equation}
Here, the spectral density of the coupling strength have been introduced
$\Gamma_{\mathrm{n}}(\omega):=2\pi\sum_{\mathbf{k}}|\mathrm{t}_{\mathrm{n},\mathbf{k}}|^{2}\delta(\omega-\epsilon_{\mathrm{n},\mathbf{k}})\simeq\Gamma_{\mathrm{n}}$
in wide-band approximation. When the electron reservoir connecting
with site-$\mathrm{n}$ is metal lead, i.e. $\Delta_{\mathrm{n}}=0$,
the metal lead only brings the dissipation $\mathsf{V}_{\mathrm{n}}(\omega)=0$
and $\mathsf{D}_{\mathrm{n}}(\omega)=\frac{1}{2}\mathsf{\Gamma}_{\mathrm{n}}\equiv\frac{1}{2}(\mathsf{\Gamma}_{\mathrm{n}}^{+}+\mathsf{\Gamma}_{\mathrm{n}}^{-})$
with $\mathsf{\Gamma}_{\mathrm{n}}^{+}=\Gamma_{\mathrm{n}}\text{diag}\{1,1,0,0\}$
and $\mathsf{\Gamma}_{\mathrm{n}}^{-}=\Gamma_{\mathrm{n}}\text{diag}\{0,0,1,1\}$
respectively. 

Then, the current $I_{\mathrm{n}}(\omega)$ in Fourier space by (\ref{eq:current-t})
is obtained as
\begin{equation}
I_{\mathrm{n}}(\omega)=-\frac{2e}{\hbar}\int\frac{d\nu}{2\pi}\sum_{\mathbf{k}}\langle\mathbf{d}^{\dagger}(\nu)\cdot\mathbf{P}^{+}\cdot\mathbf{T}_{\mathrm{n,}\mathbf{k}}\cdot\mathbf{c}_{\mathrm{n},\mathbf{k}}(\omega+\nu)\rangle.\label{eq:Fourier-current}
\end{equation}
Together with the vector operators $\mathbf{c}_{\mathbf{\mathrm{n,}\mathbf{k}}}(\omega)$
and $\mathbf{d}(\omega)$ in Eqs. (\ref{eq:lead-G}, \ref{eq:system-G}),
the current (\ref{eq:Fourier-current}) is further simplified as
\begin{align}
I_{\mathrm{n}}(\omega) & =\frac{2ie}{h}\int d\nu\{\text{tr}[\mathbf{G}^{\dagger}(\nu)\cdot\mathbf{P}^{+}\cdot\mathbb{C}_{\mathrm{n}}(\nu,\omega+\nu)-\sum_{\mathrm{m}}\mathbf{G}^{\dagger}(\nu)\cdot\mathbf{P}^{+}\cdot\mathbf{D}_{\mathrm{n}}(\omega+\nu)\cdot\mathbf{G}(\nu+\omega)\cdot\mathbb{C}_{\mathrm{m}}(\nu,\omega+\nu)]\}.\label{eq:Fourier-current-1}
\end{align}
Here, the correlation matrices of the random force are defined as
$[\mathbb{C}_{\mathrm{n}}(\nu,\omega+\nu)]_{\mathrm{ij}}:=\big\langle[\boldsymbol{\xi}_{\mathrm{n}}^{\dagger}(\nu)]_{\mathrm{j}}[\boldsymbol{\xi}_{\mathrm{n}}(\omega+\nu)]_{\mathrm{i}}\big\rangle$,
by which the relation like fluctuation-dissipation theorem is given
as \citep{Qiao_2022}
\begin{align}
\lim_{\omega\rightarrow0}\big[(-i\omega)\mathbb{C}_{\mathrm{n}}(\nu,\omega+\nu)\big] & =f_{\mathrm{n}}(\nu)\boldsymbol{\Gamma}_{\mathrm{n}}^{+}(\nu)+\bar{f}_{\mathrm{n}}(-\nu)\boldsymbol{\Gamma}_{\mathrm{n}}^{-}(\nu),\label{eq:disspation-fluctuation}
\end{align}
Here, $f_{\mathrm{n}}(\nu)$ is the electron Fermi distribution in
the initial state and $\bar{f}_{\mathrm{n}}(\nu)=1-f_{\mathrm{n}}(\nu)$
is hole distribution, respectively. Then, after a long time relaxation
and by the final value theorem $\lim_{t\rightarrow\infty}I_{\mathrm{n}}(t)=\Im\{\lim_{\omega\rightarrow0}(-i\omega)I_{\mathrm{n}}(\omega)\}$,
the generalized Landauer formula is obtained as $I_{\mathrm{n}}:=I_{\mathrm{n},\mathrm{t}}+I_{\mathrm{n},\mathrm{p}}$,
which includes the usual transport current $I_{\mathrm{n},\mathrm{t}}$
from the lead-$\mathrm{n}$ to the lead-$\mathrm{m}$:
\begin{equation}
\begin{aligned}I_{\mathrm{n},\mathrm{t}} & =\frac{e}{2\hbar}\sum_{\mathrm{m}}\int_{-\infty}^{+\infty}\frac{d\nu}{2\pi}\,\text{tr}[\mathbf{G}^{\dagger}\{\mathbf{P}^{+},\boldsymbol{\Gamma}_{\mathrm{n}}^{+}\}_{+}\mathbf{G}\boldsymbol{\Gamma}_{\mathrm{m}}^{+}]_{(\nu)}[f_{\mathrm{n}}(\nu)-f_{\mathrm{m}}(\nu)]+\text{tr}[\mathbf{G}^{\dagger}\{\mathbf{P}^{+},\boldsymbol{\Gamma}_{\mathrm{n}}^{+}\}_{+}\mathbf{G}\boldsymbol{\Gamma}_{\mathrm{m}}^{-}]_{(\nu)}[f_{\mathrm{n}}(\nu)-\bar{f}_{\mathrm{m}}(-\nu)]\\
 & +\text{tr}[\mathbf{G}^{\dagger}\{\mathbf{P}^{+},\boldsymbol{\Gamma}_{\mathrm{n}}^{-}\}_{+}\mathbf{G}\boldsymbol{\Gamma}_{\mathrm{m}}^{+}]_{(\nu)}[\bar{f}_{\mathrm{n}}(-\nu)-f_{\mathrm{m}}(\nu)]+\text{tr}[\mathbf{G}^{\dagger}\{\mathbf{P}^{+},\boldsymbol{\Gamma}_{\mathrm{n}}^{-}\}_{+}\mathbf{G}\boldsymbol{\Gamma}_{\mathrm{m}}^{-}]_{(\nu)}[\bar{f}_{\mathrm{n}}(-\nu)-\bar{f}_{\mathrm{m}}(-\nu)],
\end{aligned}
\label{eq:transport-current}
\end{equation}
and the proximity current $I_{\mathrm{n},\mathrm{p}}$ caused by the
superconducting proximity effect: 
\begin{equation}
\begin{aligned}I_{\mathrm{n},\mathrm{p}} & =i\frac{e}{\hbar}\int_{-\infty}^{+\infty}\frac{d\nu}{2\pi}\,\text{tr}\{\mathbf{G}^{\dagger}[\boldsymbol{\Gamma}_{\mathrm{n}}^{+},\mathbf{P}^{+}]_{-}\mathbf{G}[\nu-(\mathbf{H}+\sum_{\mathrm{m}}\mathbf{V}_{\mathrm{m}})]\}_{(\nu)}f_{\mathrm{n}}(\nu)+\text{tr}\{\mathbf{G}^{\dagger}[\boldsymbol{\Gamma}_{\mathrm{n}}^{-},\mathbf{P}^{+}]_{-}\mathbf{G}[\nu-(\mathbf{H}+\sum_{\mathrm{m}}\mathbf{V}_{\mathrm{m}})]\}_{(\nu)}\bar{f}_{\mathrm{n}}(-\nu)\\
 & +\sum_{\mathrm{m}}\text{tr}\{\mathbf{G}^{\dagger}[\mathbf{V}_{\mathrm{n}},\mathbf{P}^{+}]_{-}\mathbf{G}\boldsymbol{\Gamma}_{\mathrm{m}}^{+}\}_{(\nu)}f_{\mathrm{\mathrm{m}}}(\nu)+\sum_{\mathrm{m}}\text{tr}\{\mathbf{G}^{\dagger}[\mathbf{V}_{\mathrm{n}},\mathbf{P}^{+}]_{-}\mathbf{G}\boldsymbol{\Gamma}_{\mathrm{m}}^{-}\}_{(\nu)}\bar{f}_{\mathrm{m}}(-\nu).
\end{aligned}
\label{eq:proximity-current}
\end{equation}

In the above current formula (\ref{eq:transport-current}), the four
terms indicate that the electron or hole from the lead-$\mathrm{n}$
flows to the lead-$\mathrm{m}$, in which the electron or holes transforms
into electron or holes. Moreover, the four terms of (\ref{eq:proximity-current})
contribute to the average current due to the virtual process of electron
exchange brought by the superconducting proximity effect. 

When the site-$\mathrm{n}$ is connected by metal leads $\alpha=e$,
the projection operation $\mathbf{P}^{+}$ is commutative to the diagonal
dissipation matrix $[\mathbf{D}_{\mathrm{n}}(\nu),\mathbf{P}^{+}]_{-}=0$,
and no effective proximity interaction caused by the normal electrode
for system exist $\mathbf{V}_{\mathrm{n}}(\nu)=0$. Therefore, the
proximity current will not contribute to the total current, i.e.,
$I_{\mathrm{n},\mathrm{p}}=0$ and the transport current is simplified
as
\begin{equation}
I_{\mathrm{n}}=\frac{e}{h}\sum_{\mathrm{m}}\int d\nu\:\text{tr}[\mathbf{G}^{\dagger}\boldsymbol{\Gamma}_{\mathrm{n}}^{+}\mathbf{G}\boldsymbol{\Gamma}_{\mathrm{m}}^{+}]_{(\nu)}[f_{\mathrm{\mathrm{n}}}(\nu)-f_{\mathrm{\mathrm{m}}}(\nu)]+\text{tr}[\mathbf{G}^{\dagger}\boldsymbol{\Gamma}_{\mathrm{n}}^{+}\mathbf{G}\boldsymbol{\Gamma}_{\mathrm{m}}^{-}]_{(\nu)}[f_{\mathrm{n}}(\nu)-\bar{f}_{\mathrm{m}}(-\nu)].\label{eq:Landuar-formular}
\end{equation}
This simplified transport current (\ref{eq:Landuar-formular}) is
a generalized form of the Landauer formula \citep{Meir_1992_Landauer_formula,datta_1997_electronic_transport},
which has been used successfully for the current measurement caused
by the edge state in the Kitaev model \citep{Li_2014} and the nanowire-superconductor
system \citep{Qiao_2022}.

As an illustration for applying the generalized Landauer formula (\ref{eq:transport-current},
\ref{eq:proximity-current}), we calculate the supercurrent of the
light-controlled Josephson effect in Sec. \ref{sec:DcJE-Hamiltonian}
and light-controlled superconducting quantum interference devices
in Sec. \ref{sec:current-SQUID}. 

\section{Direct current Josephson effect induced by the coherence light\label{sec:DcJE-Hamiltonian} }

In this section, we consider the Hamiltonian of the two \textsl{s}-wave
superconductors, which is linked by the two-level quantum dot (QD)
driven by the classical light.
\begin{align}
H_{tot} & =H_{0}+V_{\mathrm{d}}+H_{\mathrm{L}}+H_{\mathrm{R}}+H_{t}.\label{eq:total-H-light}
\end{align}
Here, the Hamiltonian of the QD with the two-level is 
\begin{equation}
H_{0}=\Omega_{\mathrm{e}}\sum_{\sigma}d_{\mathrm{e}\sigma}^{\dagger}d_{\mathrm{e}\sigma}+\Omega_{\mathrm{g}}\sum_{\sigma}d_{\mathrm{g}\sigma}^{\dagger}d_{\mathrm{g}\sigma}.\label{eq:qdot-H}
\end{equation}
The single mode classical light drives electrons from the lower energy
level $\Omega_{\mathrm{g}}$ to the upper energy level $\Omega_{\mathrm{e}}$,
which is described by 
\begin{equation}
V_{\mathrm{d}}=-\mathcal{E}_{\mathrm{d}}d_{\mathrm{e}\sigma}^{\dagger}d_{\mathrm{g}\sigma}e^{-i\omega_{\mathrm{d}}t-i\phi_{\mathrm{d}}}-\mathcal{E}_{\mathrm{d}}d_{\mathrm{g}\sigma}^{\dagger}d_{\mathrm{e}\sigma}e^{i\omega_{\mathrm{d}}t+i\phi_{\mathrm{d}}}\label{eq:light-coupling}
\end{equation}
with the frequency $\omega_{\mathrm{d}}$ and the phase $\phi_{\mathrm{d}}$
of the driving light. And the real driving strength is defined as
$\mathcal{E}_{\mathrm{d}}\equiv\boldsymbol{E}_{\mathrm{d}}\cdot\wp_{\mathrm{eg}}$
by the transition dipole moment of quantum dot $\wp_{\mathrm{eg}}$.
The two superconducting leads are described by BCS Hamiltonian \citep{Levy_1996_Hamiltonian_approach_to_transport,Lin_2000_Electron_transport_through_mesoscopic_hybrid}
\begin{equation}
\begin{aligned}H_{\mathrm{L}} & =\sum_{\mathbf{k}}(\epsilon_{\mathrm{L},\mathbf{k}}+\mu_{\mathrm{L}})(c_{\mathrm{L},\mathbf{k}\uparrow}^{\dagger}c_{\mathrm{L},\mathbf{k}\uparrow}+c_{\mathrm{L},\mathbf{k}\downarrow}^{\dagger}c_{\mathrm{L},\mathbf{k}\downarrow})+e^{-2i\mu_{\mathrm{L}}t+i\varphi_{\mathrm{L}}}\Delta_{\mathrm{L}}c_{\mathrm{L},\mathbf{k}\uparrow}^{\dagger}c_{\mathrm{L},-\mathbf{k}\downarrow}^{\dagger}+e^{2i\mu_{\mathrm{L}}t-i\varphi_{\mathrm{L}}}\Delta_{\mathrm{L}}c_{\mathrm{L},-\mathbf{k}\downarrow}c_{\mathrm{L},\mathbf{k}\uparrow},\\
H_{\mathrm{R}} & =\sum_{\mathbf{k}}(\epsilon_{\mathrm{R},\mathbf{k}}+\mu_{\mathrm{R}})(c_{\mathrm{R},\mathbf{k}\uparrow}^{\dagger}c_{\mathrm{R},\mathbf{k}\uparrow}+c_{\mathrm{R},\mathbf{k}\downarrow}^{\dagger}c_{\mathrm{R},\mathbf{k}\downarrow})+e^{-2i\mu_{\mathrm{R}}t+i\varphi_{\mathrm{R}}}\Delta_{\mathrm{R}}c_{\mathrm{R},\mathbf{k}\uparrow}^{\dagger}c_{\mathrm{R},-\mathbf{k}\downarrow}^{\dagger}+e^{2i\mu_{\mathrm{R}}t-i\varphi_{\mathrm{R}}}\Delta_{\mathrm{R}}c_{\mathrm{R},-\mathbf{k}\downarrow}c_{\mathrm{R},\mathrm{k}\uparrow}.
\end{aligned}
\label{eq:lead-L-R}
\end{equation}
Here, the chemical potential and the phase of the superconducting
leads are $\varphi_{\mathrm{n}}$ and $\mu_{\mathrm{n}}$ with $n=\mathrm{L,R}$
respectively. And, the tunneling interaction between QD and the superconducting
leads are
\begin{equation}
H_{t}=-\sum_{\mathrm{i}=\mathrm{g,e}}\sum_{\mathrm{n},\mathbf{k},\sigma}[\mathrm{t}_{\mathrm{n,i},\mathbf{k}}d_{\mathrm{i}\sigma}^{\dagger}c_{\mathrm{n},\mathbf{k}\sigma}+\mathrm{t}_{\mathrm{n,i},\mathbf{k}}^{*}c_{\mathrm{n},\mathbf{k}\sigma}^{\dagger}d_{\mathrm{i}\sigma}].\label{eq:tunneling-L-Q-R-1}
\end{equation}
with the tunneling strength between QD and the superconducting lead
$\mathrm{t}_{\mathrm{n,i},\mathbf{k}}$. 

Then, by unitary transformation $\mathbf{S}(t)=\exp[i\{H_{0}+\sum_{\mathrm{n=L,R}}(H_{\mathrm{n}}^{s}+\mu_{\mathrm{n}}N_{\mathrm{n}})\}t]$,
where the Hamiltonian of \textit{s}-wave superconductor $H_{\mathrm{n}}^{s}$
have been defined in Eq. (\ref{eq:electron-lead}), the transformed
Hamiltonian becomes respectively 
\begin{align}
\bar{H}(t) & =\mathbf{S}(t)H_{tot}(t)\mathbf{S}^{\dagger}(t)-i\hbar\mathbf{S}(t)\frac{\partial}{\partial t}\mathbf{S}^{\dagger}(t):=\bar{H}_{\mathrm{d}}(t)+\bar{H}_{t},\label{eq:Transformed-HH}
\end{align}
where the Hamiltonian of QD driven by the coherence light is 

\begin{equation}
\bar{H}_{\mathrm{d}}(t)=-\mathcal{E}_{\mathrm{d}}\sum_{\sigma}d_{\mathrm{e}\sigma}^{\dagger}d_{\mathrm{g}\sigma}e^{i[\Omega_{\mathrm{e}}-\Omega_{\mathrm{g}}-\omega_{\mathrm{d}}]t-i\phi_{\mathrm{d}}}-\mathcal{E}_{\mathrm{d}}\sum_{\sigma}d_{\mathrm{g}\sigma}^{\dagger}d_{\mathrm{e}\sigma}e^{i[\Omega_{\mathrm{e}}-\Omega_{\mathrm{g}}-\omega_{\mathrm{d}}]t+i\phi_{\mathrm{d}}},
\end{equation}
and the tunneling Hamiltonian between QD and the superconducting leads
become
\begin{equation}
\bar{H}_{t}=-\sum_{\mathrm{n}\in\mathrm{L,R}}\sum_{\mathrm{i}\in\mathrm{g},\mathrm{e}}\sum_{\mathbf{k}\sigma}\left[\mathrm{t}_{\mathrm{n,i},\mathbf{k}}d_{\mathrm{i}\sigma}^{\dagger}e^{iH_{\mathrm{n}}^{s}t}c_{\mathrm{n},\mathbf{k}\sigma}e^{-iH_{\mathrm{n}}^{s}t}\mathrm{e}^{i(\Omega_{i}-\mu_{\mathrm{n}})t}+\mathrm{t}_{\mathrm{n,i},\mathbf{k}}^{*}e^{iH_{\mathrm{n}}^{s}t}c_{\mathrm{n},\mathbf{k}\sigma}^{\dagger}e^{-iH_{\mathrm{n}}^{s}t}\mathrm{e}^{-i(\Omega_{i}-\mu_{\mathrm{n}})t}d_{\mathrm{i}\sigma}\right].\label{eq:transfom-tunneling}
\end{equation}
Considering that the quasi-excitation in a superconductor is a mixed
excitation of electrons and holes for the superconducting lead-$\mathrm{n}$,
thus, we can utilize the inverse Bogoliubov transformation: 
\begin{equation}
\begin{aligned}c_{\mathrm{n},\mathbf{k}\uparrow} & =u_{\mathrm{n},\mathbf{k}}^{*}\alpha_{\mathrm{n},\mathbf{k}\uparrow}+v_{\mathrm{n},\mathbf{k}}\alpha_{\mathrm{n},-\mathbf{k}\downarrow}^{\dagger}\\
c_{\mathrm{n},-\mathbf{k}\downarrow}^{\dagger} & =-v_{\mathrm{n},\mathbf{k}}^{*}\alpha_{\mathrm{n},\mathbf{k}\uparrow}+u_{\mathrm{n},\mathbf{k}}\alpha_{\mathrm{n},-\mathbf{k}\downarrow}^{\dagger}
\end{aligned}
\label{eq:Bogoliubov_tran}
\end{equation}
with $u_{\mathrm{n},\mathbf{k}}=\sqrt{(1+\epsilon_{\mathrm{n},\mathbf{k}}/E_{\mathrm{n},\mathbf{k}})/2}$
and $v_{\mathrm{n},\mathbf{k}}=e^{i\varphi_{\mathrm{n}}}\sqrt{(1-\epsilon_{\mathrm{n},\mathbf{k}}/E_{\mathrm{n},\mathbf{k}})/2}$,
to simplify the following relation 
\begin{equation}
\begin{aligned}e^{iH_{\mathrm{n}}^{s}t}c_{\mathrm{n},\mathbf{k}\sigma}e^{-iH_{\mathrm{n}}^{s}t} & =u_{\mathrm{n},\mathbf{k}}^{*}\alpha_{\mathrm{n},\mathbf{k}\uparrow}e^{-iE_{\mathrm{n},\mathbf{k}}t}+v_{\mathrm{n},\mathbf{k}}\alpha_{\mathrm{n},-\mathbf{k}\downarrow}^{\dagger}e^{iE_{\mathrm{n},\mathbf{k}}t},\\
e^{iH_{\mathrm{n}}^{s}t}c_{\mathrm{n},-\mathbf{k}\downarrow}^{\dagger}e^{-iH_{\mathrm{n}}^{s}t} & =-v_{\mathrm{n},\mathbf{k}}^{*}\alpha_{\mathrm{n},\mathbf{k}\uparrow}e^{-iE_{\mathrm{n},\mathbf{k}}t}+u_{\mathrm{n},\mathbf{k}}\alpha_{\mathrm{n},-\mathbf{k}\downarrow}^{\dagger}e^{iE_{\mathrm{n},\mathbf{k}}t}.
\end{aligned}
\label{eq:exication_evolution}
\end{equation}
Here, the superconducting quasi-excitation energy is $E_{\mathrm{n},\mathbf{k}}=\sqrt{\epsilon_{\mathrm{n},\mathbf{k}}^{2}+|\Delta_{\mathrm{n}}|^{2}}$. 

It follows from Eqs. (\ref{eq:transfom-tunneling}, \ref{eq:exication_evolution})
that the time oscillation of the tunneling strength between the lead-L
and the lower level is much smaller than the coupling tunneling strength
between the lead-L and the upper level in such case: $|\Omega_{\mathrm{g}}-\mu_{\mathrm{L}}\pm E_{\mathrm{n},\mathbf{k}}|\ll|\Omega_{\mathrm{e}}-\mu_{\mathrm{L}}\pm E_{\mathrm{n},\mathbf{k}}|$.
Therefore, we can ignore the coupling tunneling between the upper
level and the lead-L under the rotating wave approximation, that is,
$\mathrm{t}_{\mathrm{L,g},\mathbf{k}}e^{i(\Omega_{\mathrm{g}}-\mu_{\mathrm{L}}\pm E_{\mathrm{n},\mathbf{k}})}\gg\mathrm{t}_{\mathrm{L,e},\mathbf{k}}e^{i(\Omega_{\mathrm{e}}-\mu_{\mathrm{L}}\pm E_{\mathrm{n},\mathbf{k}})}\simeq0$,
as shown in Fig. \ref{Fig:effective-system} (a). Similarly, we can
also ignore the coupling tunneling between the lower level and the
the lead-R $\mathrm{t}_{\mathrm{R,g},\mathbf{k}}\simeq0$. So far,
we have proven that the left and right superconducting lead, respectively
coupled to the low level $\Omega_{\mathrm{g}}$ and upper level $\Omega_{\mathrm{e}}$
is reasonable under the rotating wave approximation, as considered
in the main text.

Then, we can calculate the superconducting current of the light-controlled
Josephson effect in the rotating frame with respect to the unitary
transformation 
\begin{equation}
\mathbf{U}(t)\equiv\mathbf{S}(t)=\exp[i(\hat{N}_{\mathrm{L}}+\sum_{\sigma}d_{\mathrm{g}\sigma}^{\dagger}d_{\mathrm{g}\sigma})\mu_{\mathrm{L}}t+i(\hat{N}_{\mathrm{R}}+\sum_{\sigma}d_{\mathrm{e}\sigma}^{\dagger}d_{\mathrm{e}\sigma})(\mu_{\mathrm{R}}t+\phi_{\mathrm{d}})].\label{eq:U_transform}
\end{equation}
According to Eq. (\ref{eq:Transformed-HH}), the specific form of
the Hamiltonian of light-driven QD, superconducting lead and tunneling
interaction are respectively 
\begin{equation}
\begin{aligned}\bar{H}_{\mathrm{QD}} & =\bar{\Omega}_{\mathrm{e}}\sum_{\sigma}d_{\mathrm{e}\sigma}^{\dagger}d_{\mathrm{e}\sigma}+\bar{\Omega}_{\mathrm{g}}\sum_{\sigma}d_{\mathrm{g}\sigma}^{\dagger}d_{\mathrm{g}\sigma}-\sum_{\sigma}[\mathcal{E}_{\mathrm{d}}d_{\mathrm{e}\sigma}^{\dagger}d_{\mathrm{g}\sigma}e^{i(\mu_{\mathrm{R}}-\mu_{\mathrm{L}}-\omega_{\mathrm{d}})t}+\mathcal{E}_{\mathrm{d}}d_{\mathrm{g}\sigma}^{\dagger}d_{\mathrm{e}\sigma}e^{-i(\mu_{\mathrm{R}}-\mu_{\mathrm{L}}-\omega_{\mathrm{d}})t}],\\
H_{\mathrm{L}} & =\sum_{\mathbf{k}}\epsilon_{\mathrm{L},\mathbf{k}}(c_{\mathrm{L},\mathbf{k}\uparrow}^{\dagger}c_{\mathrm{L},\mathbf{k}\uparrow}+c_{\mathrm{L},\mathbf{k}\downarrow}^{\dagger}c_{\mathrm{L},\mathbf{k}\downarrow})+e^{i\varphi_{\mathrm{L}}}\Delta_{\mathrm{L}}c_{\mathrm{L},\mathbf{k}\uparrow}^{\dagger}c_{\mathrm{L},-\mathbf{k}\downarrow}^{\dagger}+e^{-i\varphi_{\mathrm{L}}}\Delta_{\mathrm{L}}c_{\mathrm{L},-\mathbf{k}\downarrow}c_{\mathrm{L},\mathbf{k}\uparrow},\\
H_{\mathrm{R}} & =\sum_{\mathbf{k}}\epsilon_{\mathrm{R},\mathbf{k}}(c_{\mathrm{R},\mathbf{k}\uparrow}^{\dagger}c_{\mathrm{R},\mathbf{k}\uparrow}+c_{\mathrm{R},\mathbf{k}\downarrow}^{\dagger}c_{\mathrm{R},\mathbf{k}\downarrow})+e^{i\bar{\varphi}_{\mathrm{R}}}\Delta_{\mathrm{R}}c_{\mathrm{R},\mathbf{k}\uparrow}^{\dagger}c_{\mathrm{R},-\mathbf{k}\downarrow}^{\dagger}+e^{-i\bar{\varphi}_{\mathrm{R}}}\Delta_{\mathrm{R}}c_{\mathrm{R},-\mathbf{k}\downarrow}c_{\mathrm{R},\mathrm{k}\uparrow}\\
H_{t} & =-\sum_{\mathbf{k},\sigma}[\mathrm{t}_{\mathrm{R},\mathrm{e},\mathbf{k}}d_{\mathrm{e}\sigma}^{\dagger}c_{\mathrm{R},\mathbf{k}\sigma}+\mathrm{t}_{\mathrm{L},\mathrm{g},\mathbf{k}}d_{\mathrm{g}\sigma}^{\dagger}c_{\mathrm{L},\mathbf{k}\sigma}+\mathrm{H.c.}].
\end{aligned}
\label{eq:U-H}
\end{equation}
Here, the reduced upper and lower level are defined as $\bar{\Omega}_{\mathrm{e}}:=\Omega_{\mathrm{e}}-\mu_{\mathrm{R}}$
and $\bar{\Omega}_{\mathrm{g}}:=\Omega_{\mathrm{g}}-\mu_{\mathrm{L}}$.
And it is seen from the Hamiltonian of the right superconducting lead
in the third row of Eq. (\ref{eq:U-H}) that the phase of superconducting
lead-R become $\bar{\varphi}_{\mathrm{R}}=\varphi_{\mathrm{R}}+2\phi_{\mathrm{d}}$
in comparison with (\ref{eq:lead-L-R}). Moreover, the energy level
of the light-controlled quantum dot connected by two superconducting
leads from Fig. \ref{Fig:effective-system} (a) effectively become
Fig. \ref{Fig:effective-system} (b) in the rotated representation. 

When the frequency of the driving light is equal to the difference
between the chemical potentials of the left and right superconducting
leads, i.e. $\mu_{\mathrm{R}}-\mu_{\mathrm{L}}=\omega_{\mathrm{d}}$,
the Hamiltonian of the total system becomes time-independent. Then,
we rewrite the Hamiltonian of the light-driven QD, superconducting
leads, and tunneling interaction as the form of Eqs (\ref{eq:measured-system},
\ref{eq:electron-lead}, \ref{eq:tunneling})
\begin{align}
\bar{H}_{\mathrm{QD}} & =\frac{1}{2}\mathbf{d}^{\dagger}\cdot\mathbf{H}\cdot\mathbf{d}\equiv\frac{1}{2}\mathbf{d}^{\dagger}\cdot\begin{bmatrix}\bar{\Omega}_{\mathrm{e}} & 0 & 0 & 0 & -\mathcal{E}_{\mathrm{d}} & 0 & 0 & 0\\
0 & \bar{\Omega}_{\mathrm{e}} & 0 & 0 & 0 & -\mathcal{E}_{\mathrm{d}} & 0 & 0\\
0 & 0 & -\bar{\Omega}_{\mathrm{e}} & 0 & 0 & 0 & \mathcal{E}_{\mathrm{d}} & 0\\
0 & 0 & 0 & -\bar{\Omega}_{\mathrm{e}} & 0 & 0 & 0 & \mathcal{E}_{\mathrm{d}}\\
-\mathcal{E}_{\mathrm{d}} & 0 & 0 & 0 & \bar{\Omega}_{\mathrm{g}} & 0 & 0 & 0\\
0 & -\mathcal{E}_{\mathrm{d}} & 0 & 0 & 0 & \bar{\Omega}_{\mathrm{g}} & 0 & 0\\
0 & 0 & \mathcal{E}_{\mathrm{d}} & 0 & 0 & 0 & -\bar{\Omega}_{\mathrm{g}} & 0\\
0 & 0 & 0 & \mathcal{E}_{\mathrm{d}} & 0 & 0 & 0 & -\bar{\Omega}_{\mathrm{g}}
\end{bmatrix}\cdot\mathbf{d},\label{eq:H}\\
H_{\mathrm{\mathrm{n}}}^{\mathrm{s}} & =\frac{1}{2}\mathbf{c}_{\mathrm{n},\mathbf{k}}^{\dagger}\cdot\mathbf{H}_{\mathrm{n}}\cdot\mathbf{c}_{\mathrm{n},\mathbf{k}}\equiv\frac{1}{2}\mathbf{c}_{\mathrm{n},\mathbf{k}}^{\dagger}\cdot\begin{bmatrix}\epsilon_{\mathrm{\mathrm{n}},\mathbf{k}} &  &  & \Delta_{\mathrm{n}}\\
 & \epsilon_{\mathrm{\mathrm{n}},\mathbf{k}} & -\Delta_{\mathrm{n}}\\
 & -\Delta_{\mathrm{n}}^{*} & -\epsilon_{\mathrm{\mathrm{n}},\mathbf{k}}\\
\Delta_{\mathrm{\mathrm{n}}}^{*} &  &  & -\epsilon_{\mathrm{\mathrm{n}},\mathbf{k}}
\end{bmatrix}\cdot\mathbf{c}_{\mathrm{n},\mathbf{k}},\mathrm{n=L,R},\label{eq:lead-s}\\
H_{t} & =-\sum_{\mathbf{k},\sigma}[\mathrm{t}_{\mathrm{R},\mathrm{e},\mathbf{k}}d_{\mathrm{e}\sigma}^{\dagger}c_{\mathrm{R},\mathbf{k}\sigma}+\mathrm{t}_{\mathrm{L},\mathrm{g},\mathbf{k}}d_{\mathrm{g}\sigma}^{\dagger}c_{\mathrm{L},\mathbf{k}\sigma}+\mathrm{H.c.}].\label{eq:tunneling-1}
\end{align}
where the vector operator is $\mathbf{d}=[d_{\mathrm{e}\uparrow},d_{\mathrm{e}\downarrow},d_{\mathrm{e}\uparrow}^{\dagger},d_{\mathrm{e}\downarrow}^{\dagger},d_{\mathrm{g}\uparrow},d_{\mathrm{g}\downarrow},d_{\mathrm{g}\uparrow}^{\dagger},d_{\mathrm{g}\downarrow}^{\dagger}]^{T}$
for $\mathrm{n=e,g}$ and the pairing strength of the left and right
superconducting lead are $\Delta_{\mathrm{L}}=\Delta_{\mathrm{s}}e^{i\varphi_{\mathrm{L}}}$
and $\Delta_{\mathrm{R}}=\Delta_{\mathrm{s}}e^{i(\varphi_{\mathrm{R}}+2\phi_{\mathrm{d}})}$.
Notice that the superconducting phase of the pairing strength for
the right superconducting lead can be adjusted by the phase of coherence
light $\phi_{\mathrm{d}}$.

\begin{figure}
\includegraphics[width=1\columnwidth]{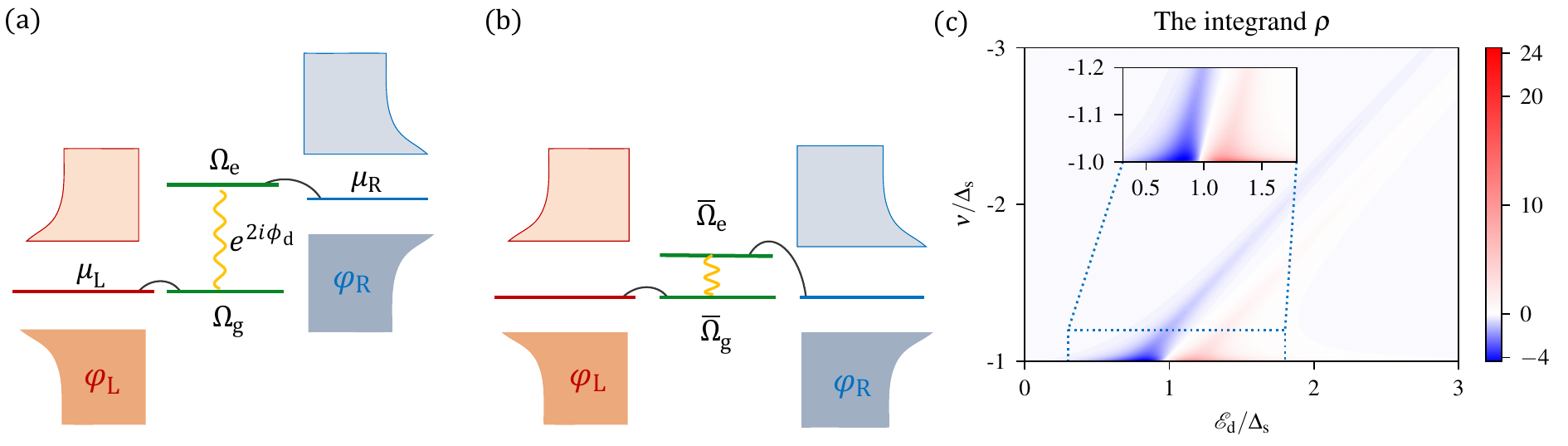}

\caption{The energy level (a) of the two superconducting leads connected by
a light-controlled quantum dot in Eq. (\ref{eq:total-H-light}) by
the unitary transformation $\mathbf{U}(t)$ becomes the effective
energy level (b) with the reduced lower level $\bar{\Omega}_{\mathrm{g}}=\Omega_{\mathrm{g}}-\mu_{\mathrm{L}}$and
upper level $\bar{\Omega}_{\mathrm{e}}=\Omega_{\mathrm{e}}-\mu_{\mathrm{R}}$,
and the phase $\bar{\varphi}_{\mathrm{R}}=\varphi_{\mathrm{R}}+2\phi_{\mathrm{d}}$
of the right superconducting of (\ref{eq:U-H}) in the rotation representation.
Plotted (c) shows that the main contribution of the integrand $\rho(\nu)$
in Eq. (\ref{eq:intergrand}) to the integral is concentrated between
$-1.2\Delta_{\mathrm{s}}$ and $-\Delta_{\mathrm{s}}$ when the driving
strength $\mathcal{E}_{\mathrm{d}}$ changes from $0.3\Delta_{\mathrm{s}}$
to $1.8\Delta_{\mathrm{s}}$, which is shown more clearly in the inset.}

\label{Fig:effective-system}
\end{figure}

Thus, according to Eq. (\ref{eq:transport-current}, \ref{eq:proximity-current}),
the transport current (\ref{eq:transport-current}) is zero due to
the same chemical potentials of the left and right superconducting
lead, and the proximity current (\ref{eq:proximity-current}) induced
by the light simplified as
\begin{equation}
I_{\mathrm{s}}=\frac{e}{2\hbar}\int_{-\infty}^{+\infty}\frac{d\nu}{2\pi}\,\text{tr}\{[\mathbf{G}(\nu)-\mathbf{G}^{\dagger}(\nu)][\boldsymbol{\Gamma}_{\mathrm{L}}^{+}(\nu),\mathbf{P}^{+}]_{-}\}f_{\mathrm{L}}(\nu)++\text{tr}\{[\mathbf{G}(\nu)-\mathbf{G}^{\dagger}(\nu)][\boldsymbol{\Gamma}_{\mathrm{L}}^{-}(\nu),\mathbf{P}^{+}]_{-}\}\bar{f}_{\mathrm{L}}(-\nu).\label{eq:simplied-current}
\end{equation}
In deriving above current formula, the following relation is needed
\begin{equation}
\mathbf{G}^{\dagger}(\nu)-\mathbf{G}(\nu)=-2i\mathbf{G}(\nu)(\nu-[\mathbf{H}+\sum_{\mathrm{m}}\mathbf{V}_{\mathrm{m}}(\nu)])\mathbf{G}^{\dagger}(\nu).\label{eq:G-G}
\end{equation}
Together with the Hamiltonian $\mathbf{H}$ of the measured system
(\ref{eq:H}) and the dissipation kernel $\mathbf{D}(\omega)=\sum_{\mathrm{m=L,R}}\mathbf{D}_{\mathrm{m}}(\omega)$
caused by the two superconducting leads in Eq. (\ref{eq:disspation}),
the Green function $\mathbf{G}(\omega)=i[\omega^{+}-\mathbf{H}+i\mathbf{D}(\omega)]^{-1}$
is obtained. Further, the dc current induced by the light, which is
completely driven by the superconducting phase difference $\Phi:=\bar{\varphi}_{\mathrm{R}}-\varphi_{\mathrm{L}}=2\phi_{\mathrm{d}}+\varphi_{\mathrm{R}}-\varphi_{\mathrm{L}}$,
are exactly calculated as 

\begin{align}
I_{\mathrm{s}} & =\frac{e}{h}\int d\nu\frac{256\Gamma^{3}\Delta_{\mathrm{s}}^{2}\nu^{2}\mathcal{E}_{\mathrm{d}}^{2}\left(\Omega_{\Gamma}^{2}-4\nu^{2}\right)\text{Sgn}[\nu]\Re[\gamma_{\mathrm{s}}(\nu)]f_{\mathrm{L}}(\nu)\sin(\Phi)}{64\Gamma^{2}\nu^{4}\gamma_{\mathrm{s}}^{2}(\nu)(\Omega_{\Gamma}^{2}-4\nu^{2})^{2}+\left(4\Gamma^{2}\left(\Delta_{\mathrm{s}}^{2}\Omega_{\delta}^{2}(\Phi)+\nu^{2}(4\nu^{2}-\Omega^{2})\right)-\gamma_{\mathrm{s}}^{2}(\nu)[(\Gamma^{2}-4\nu^{2})^{2}-16(\nu^{2}\Omega^{2}-\mathcal{E}_{\mathrm{d}}^{4})]\right)^{2}}\nonumber \\
 & :=I_{c}(\Phi)\sin(\Phi).\label{eq:dc-current}
\end{align}
Here, we have considered the reduced lower energy for simplicity $\bar{\Omega}_{\mathrm{g}}=\Omega_{\mathrm{g}}-\mu_{\mathrm{L}}\equiv0$
and the same coupling strength $\Gamma_{\mathrm{L}}=\Gamma_{\mathrm{R}}=\Gamma$,
and the light-atom detuning is defined as $\delta\omega=\Omega_{\mathrm{e}}-\Omega_{\mathrm{g}}-\omega_{\mathrm{d}}$.
At the same time, the following notations are also used for simplifying
the above formula
\begin{equation}
\Omega=\sqrt{\delta\omega^{2}+2\mathcal{E}_{\mathrm{d}}^{2}},\quad\Omega_{\Gamma}=\sqrt{2\Omega^{2}+\Gamma^{2}},\quad\Omega_{\delta}^{2}(\Phi)=\delta\omega^{2}+2\mathcal{E}_{\mathrm{d}}^{2}\cos\Phi,\quad\gamma_{\mathrm{s}}(\nu)=\sqrt{\nu^{2}-\Delta_{\mathrm{s}}^{2}}.
\end{equation}
In the large detuning situation ($\delta\omega\gg\mathcal{E}_{\mathrm{d}}$),
since $\Omega_{\mathrm{\delta}}^{2}(\Phi)=\delta\omega^{2}+2\mathcal{E}_{\mathrm{d}}^{2}\cos\Phi\simeq\delta\omega^{2}$,
the above critical current $I_{c}(\Phi)\simeq I_{c}$ becomes a constant
independent of the phase difference $\Phi$, then the above supercurrent
(\ref{eq:dc-current}) is simplified as $I_{\mathrm{s}}=I_{c}\sin\Phi$.
And, it can be seen from Eq. (\ref{eq:dc-current}) that the integrand
is given as 
\begin{equation}
\rho(\nu)=\frac{\Delta_{\mathrm{s}}^{2}\nu^{2}\mathcal{E}_{\mathrm{d}}^{2}\left(\Omega_{\Gamma}^{2}-4\nu^{2}\right)\text{Sgn}[\nu]\Re[\gamma_{\mathrm{s}}(\nu)]f_{\mathrm{L}}(\nu)}{64\Gamma^{2}\nu^{4}\gamma_{\mathrm{s}}^{2}(\nu)(\Omega_{\Gamma}^{2}-4\nu^{2})^{2}+\left(4\Gamma^{2}\left(\Delta_{\mathrm{s}}^{2}\Omega_{\delta}^{2}(\Phi)+\nu^{2}(4\nu^{2}-\Omega^{2})\right)-\gamma_{\mathrm{s}}^{2}(\nu)[(\Gamma^{2}-4\nu^{2})^{2}-16(\nu^{2}\Omega^{2}-\mathcal{E}_{\mathrm{d}}^{4})]\right)^{2}},\label{eq:intergrand}
\end{equation}
the main contribution of which to the integral comes from near $\nu\sim-\Delta_{\mathrm{s}}$
due to $\left(\nu^{2}-\Delta_{\mathrm{s}}^{2}\right)^{-\frac{1}{2}}\rightarrow\infty$
when $\nu\rightarrow-\Delta_{\mathrm{s}}-\varepsilon$ ($\varepsilon$
is infinitesimal), which can be proved exactly by numerical computation,
as shown in Fig. \ref{Fig:effective-system} (c). 

\section{The supercurrent through SQUID with two quantum dots\label{sec:current-SQUID} }

In this section, we consider that the two same QDs embedded in a superconducting
quantum interference devices (SQUID) loop is driven by the coherence
of light. According to the Hamiltonian of the QD coupled by the light
in Eqs (\ref{eq:qdot-H}, \ref{eq:light-coupling}), the two QDs coupled
through the coherence light is described by 
\begin{equation}
H_{s}=\sum_{i=1,2}[\Omega_{\mathrm{e}}\sum_{\sigma}d_{i,\mathrm{e}\sigma}^{\dagger}d_{i,\mathrm{e}\sigma}+\Omega_{\mathrm{g}}\sum_{\sigma}d_{i,\mathrm{g}\sigma}^{\dagger}d_{i,\mathrm{g}\sigma}-\mathcal{E}_{i,\mathrm{d}}d_{i,\mathrm{e}\sigma}^{\dagger}d_{i,\mathrm{g}\sigma}e^{-i\omega_{\mathrm{d}}t-i\phi_{i,\mathrm{d}}}-\mathcal{E}_{i,\mathrm{d}}d_{i,\mathrm{g}\sigma}^{\dagger}d_{i,\mathrm{e}\sigma}e^{i\omega_{\mathrm{d}}t+i\phi_{i,\mathrm{d}}}].\label{eq:SQUID-system}
\end{equation}
Here, we have considered the same the frequency of the driving light
$\omega_{\mathrm{d},1}=\omega_{\mathrm{d},2}=\omega_{\mathrm{d}}.$
Similarly, the tunneling interaction between QD and the superconducting
leads under the rotating wave approximation are rewritten as 
\begin{equation}
\hat{H}_{t}=-\sum_{i=1,2}\sum_{\mathbf{k},\sigma}[\mathrm{t}_{\mathrm{R},\mathrm{e},\mathbf{k}}d_{i,\mathrm{e}\sigma}^{\dagger}c_{\mathrm{R},\mathbf{k}\sigma}+\mathrm{t}_{\mathrm{L},\mathrm{g},\mathbf{k}}d_{i,\mathrm{g}\sigma}^{\dagger}c_{\mathrm{L},\mathbf{k}\sigma}+\mathrm{H.c.}].\label{eq:tunneling-L-2QD-R}
\end{equation}

In the case where the light frequency is equal to the difference of
chemical potential between two superconducting leads, i.e. $\omega_{\mathrm{d}}=\mu_{\mathrm{R}}-\mu_{\mathrm{L}}$,
we can introduce unitary transform 
\begin{equation}
\mathbf{U}(t)=\exp[i(N_{\mathrm{L}}+\sum_{i\sigma}d_{i,\mathrm{g}\sigma}^{\dagger}d_{i,\mathrm{g}\sigma})\mu_{\mathrm{L}}t+i(N_{\mathrm{R}}+\sum_{i\sigma}d_{i,\mathrm{e}\sigma}^{\dagger}d_{i,\mathrm{e}\sigma})\mu_{\mathrm{R}}t+i\sum_{i\sigma}d_{i,\mathrm{e}\sigma}^{\dagger}d_{i,\mathrm{e}\sigma})\phi_{i,\mathrm{d}}],
\end{equation}
to obtain the transformed Hamiltonian: $\bar{H}_{s}=\frac{1}{2}\mathbf{d}^{\dagger}\cdot\mathbf{H}\cdot\mathbf{d}$,
where the Hamiltonian matrix in vector operator basis $\mathbf{d}=[\mathsf{d}_{1,\mathrm{e}},\mathsf{d}_{1,\mathrm{g}},\mathsf{d}_{2,\mathrm{e}},\mathsf{d}_{2,\mathrm{g}}]^{T}$
is $\mathbf{H}=\mathrm{diag}\{\mathbf{H}_{Q,1},\mathbf{H}_{Q,2}\}$
with the block matrix $\mathbf{H}_{Q,i}$ defined in Eq. (\ref{eq:H}).
Respectively, by defining the vector operator of the right superconducting
lead
\begin{equation}
\bar{\mathbf{c}}_{i,\mathrm{R},\mathbf{k}}=[\bar{c}_{i,\mathrm{R},\mathbf{k}\uparrow},\bar{c}_{i,\mathrm{R},-\mathbf{k}\downarrow},\bar{c}_{i,\mathrm{R},\mathbf{k}\uparrow}^{\dagger},\bar{c}_{i,\mathrm{R},-\mathbf{k}\downarrow}^{\dagger}]^{T}\equiv[e^{i\phi_{i,\mathrm{d}}}c_{\mathrm{R},\mathbf{k}\uparrow},e^{i\phi_{i,\mathrm{d}}}c_{\mathrm{R},-\mathbf{k}\downarrow},e^{-i\phi_{i,\mathrm{d}}}c_{\mathrm{R},\mathbf{k}\uparrow}^{\dagger},e^{-i\phi_{i,\mathrm{d}}}c_{\mathrm{R},-\mathbf{k}\downarrow}^{\dagger}]^{T},
\end{equation}
we can obtain the tunneling Hamiltonian between QD and the superconducting
leads 
\begin{equation}
H_{t}=-\sum_{i=1,2}\sum_{\mathbf{k},\sigma}[\mathrm{t}_{\mathrm{R},\mathrm{e},\mathbf{k}}d_{i,\mathrm{e}\sigma}^{\dagger}\bar{c}_{i,\mathrm{R},\mathbf{k}\sigma}+\mathrm{t}_{\mathrm{L},\mathrm{g},\mathbf{k}}d_{i,\mathrm{g}\sigma}^{\dagger}c_{\mathrm{L},\mathbf{k}\sigma}+\mathrm{H.c.}].\label{eq:transformed-2QD-tunneling}
\end{equation}
and the Hamiltonian of the superconducting lead-$\mathrm{n}$: 
\begin{align}
\bar{H}_{i,\mathrm{n}}^{\mathrm{s}} & =\frac{1}{2}\bar{\mathbf{c}}_{i,\mathrm{n},\mathbf{k}}^{\dagger}\cdot\mathbf{H}_{i,\mathrm{n}}\cdot\bar{\mathbf{c}}_{i,\mathrm{n},\mathbf{k}}\equiv\frac{1}{2}\bar{\mathbf{c}}_{i,\mathrm{n},\mathbf{k}}^{\dagger}\cdot\begin{bmatrix}\epsilon_{\mathrm{n},\mathbf{k}} &  &  & \Delta_{i,\mathrm{n}}\\
 & \epsilon_{\mathrm{n},\mathbf{k}} & -\Delta_{i,\mathrm{n}}\\
 & -\Delta_{i,\alpha}^{*} & -\epsilon_{\mathrm{n},\mathbf{k}}\\
\Delta_{i,\mathrm{n}}^{*} &  &  & -\epsilon_{\mathrm{n},\mathbf{k}}
\end{bmatrix}\cdot\bar{\mathbf{c}}_{i,\mathrm{n},\mathbf{k}},\:\mathrm{n}=\mathrm{L,R},\label{eq:lead-s-1}
\end{align}
where the pairing term of the left and right superconducting lead
are $\Delta_{i,\mathrm{L}}=\Delta_{\mathrm{s}}e^{i\varphi_{\mathrm{L}}}$
and $\Delta_{i,\mathrm{R}}=\Delta_{\mathrm{s}}e^{i(\varphi_{\mathrm{R}}+2\phi_{i,\mathrm{d}})}$.

As the block diagonal Hamiltonian $\mathbf{H}=\mathrm{diag}\{\mathbf{H}_{Q,1},\mathbf{H}_{Q,2}\}$
of the measured system and the dissipation kenal $\mathbf{D}(\omega)=\sum_{\mathrm{m=L,R}}\mathbf{D}_{\mathrm{m}}(\omega)$
in Eq. (\ref{eq:disspation}) are block diagonal, the Green function
$\mathbf{G}(\omega)=i[\omega^{+}-\mathbf{H}+i\mathbf{D}(\omega)]^{-1}$
is also rewritten as $8\times8$ block diagonal matrix. Therefore,
also using Eq. (\ref{eq:simplied-current}), the total current is
the sum of the current flowing through two QDs respectively 
\begin{equation}
I_{\mathrm{s}}=I_{\mathrm{s},1}+I_{\mathrm{s},2}\label{eq:SQUID-current}
\end{equation}
Here, the current through each quantum dot is $I_{\mathrm{s},i}=I_{c,i}(\Phi_{i})\sin\Phi_{i}$
with total phase $\Phi_{1}=\varphi_{\mathrm{R}}-\varphi_{\mathrm{L}}+2\phi_{1,\mathrm{d}}$
and $\Phi_{2}=\varphi_{\mathrm{R}}-\varphi_{\mathrm{L}}+2\phi_{2,\mathrm{d}}$,
which is given in Eq. (\ref{eq:dc-current}). 

\end{widetext}

\bibliographystyle{apsrev4-2}
\bibliography{Refs}

\end{document}